\documentstyle[a4p,12pt,amsbsy,epsf]{article}
\newcommand {\nbsphysBG}     {\mbox{$35 \pm 9$}}
\newcommand {\ncanddsl}   {\mbox{244}} % AH 1/00 number of signal candidates
\newcommand {\nsideband}  {\mbox{199}} % AH      number of sideband candidates

\newcommand {\ncandbs}     {\mbox{116}}  % num of B_s sig cands w/o phys & BG
\newcommand {\ncandbserr}  {\mbox{10}}   % error on ncandbs.
\newcommand {\sensitivity} {4.1} % experimental sensitivity
\newcommand {\limit}       {1.0} % experimental lower limit
\newcommand {\Zzero} {\mbox{$\rm{ Z}^0$}}

\newcommand {\ee} {\mbox{$e^+e^-$}}

\newcommand {\Ecm}   {\mbox{${E_{\mathrm cm}}$}}
\newcommand {\bbar}       {\mathrm{\bar{b}}}
\newcommand {\bbbar}      {\mathrm{b\bar{b}}}

\newcommand {\qqbar}      {\mathrm{q\bar{q}}}
\newcommand {\lm}         {\ell^-}
\newcommand {\lp}         {\ell^+}

\newcommand {\nubar}      {\bar{\nu}}
\newcommand {\K}          {\mathrm{K}}
\newcommand {\D}          {\mathrm{D}}
\newcommand {\B}          {\mathrm{B}}

\newcommand {\X}          {\mathrm{(X)}}
\newcommand {\Xnop}       {\mathrm{X}}
 
\def\Kstar{\mbox{$\mathrm{ K^{*0}}$}}
\def\Km{\mbox{$\mathrm{ K^-}$}}

\newcommand {\Dsp}        {\mbox{$\mathrm{ D_s^+}$}}
\newcommand {\Dsm}        {\mbox{$\mathrm{ D_s^-}$}}
\newcommand {\Ds}         {\Dsm}

\newcommand {\Lcp}        {\mathrm{\Lambda_c^+}}
\newcommand {\Lc}         {\Lcp}  % this macro used - AH 3/10/99
\newcommand {\Bunobox}        {\mathrm{B_u}}
\newcommand {\Bdnobox}        {\mathrm{B_d}}
\newcommand {\Bud}        {\mbox{$\mathrm{ B_{(u/d)}}$}}
\newcommand {\Budbar}        {\mbox{$\mathrm{ \overline{B}_{(u/d)}}$}}

\def\Bs{\mbox{$\mathrm{B_s}$}}
\def\Bsbar{\mbox{$\mathrm{\overline{B}_s}$}}
\def\Bd{\mbox{$\mathrm{B_d}$}}
\def\Bu{\mbox{$\mathrm{B_u}$}}
\def\Bdbar{\mbox{$\mathrm{\overline{B}_d}$}}
\def\Bubar{\mbox{$\mathrm{\overline{B}_u}$}}
\newcommand {\phipi}      {\phi\pi^-}
\newcommand {\Kst}        {\K^{*0}}
\newcommand {\KstK}       {\Kst\K^-}

\newcommand {\philnu}     {\phi\lm\nubar}
\newcommand {\philnux}    {\phi\lm\nubar\X}

\newcommand {\KKpi}       {\K^+\K^-\pi^-}

\newcommand {\KKp}        {\KKpi}

\newcommand {\Ko}         {\mathrm{K^0_s}}
\newcommand {\KKo}        {\Ko\K^-}
\newcommand {\Dsl}        {\Dsm\lp}
\newcommand {\Dslwr}      {\Dsm\lm}

\newcommand {\Dslnux}     {\Dsm\lp\nu\X}
\newcommand {\ud}         {\mbox{$\mathrm{d}$}}
\newcommand {\Br}         {\mathrm{Br}}
\newcommand {\dEdx}       {\mbox{${\mathrm{d}}E/{\mathrm{d}}x$}}
\newcommand {\etal}       {{et al.}}

\newcommand {\ex}         {\cdot 10^}
\def\dms{\Delta m_{\mathrm{s}}}
\def\dmd{\Delta m_{\mathrm{d}}}

\newcommand {\dl}         {\sigma^l}
\newcommand {\dlrec}      {\dl_{\mathrm recon}}

\newcommand {\pbrec}      {p_{\mathrm recon}}
\newcommand {\pr}         {{\mathrm P}}

\newcommand{\fother}       {f_{\mathrm other}}
\newcommand{\fp}           {f_{\rm bg}^+}
\newcommand{\fplep}        {f_{\rm bg,\ts lep}^+}
\newcommand{\fmassi}       {f^{\rm{mass}}_i}
\newcommand{\fmixi}        {f^{\rm{mix}}_i}
\newcommand{\fphysi}       {f^{\rm{phys}}_i}
\newcommand{\fbgi}         {f^{\rm{comb}}_i}
\newcommand{\fphysa}       {Y^a}
\newcommand{\fphysb}       {Y^b}
\newcommand{\fraca}        {f_d^a}
\newcommand{\fracb}        {f_d^b}
\newcommand{\fOCB}         {C}

\newcommand{\fOCBdphi}     {C_0^{\phi}}
\newcommand{\fOCBsphi}     {C_s^{\phi}}
\newcommand{\fOCBdno}      {C_0^{\mathrm no~\phi}}
\newcommand{\fOCBsno}      {C_s^{\mathrm no~\phi}}

\newcommand {\taubs}      {\mbox{$\mathrm{\tau_{B_s}}$}}

\newcommand{\tbd}           {\tau_{\Bdnobox}}
\newcommand{\tbu}           {\tau_{\Bunobox}}
\newcommand{\tbgplep}         {\tau_{\mathrm bg,\ts lep}^+}
\newcommand{\tbgphad}         {\tau_{\mathrm bg,\ts had}^+}
\newcommand{\tbgnlep}         {\tau_{\mathrm bg,\ts lep}^-}
\newcommand{\ftmid}         {\alpha_{\mathrm mix}}
\newcommand{\ftend}         {\beta_{\mathrm mix}}
\newcommand{\Qsame}           {Q_{\mathrm same}}
\newcommand{\Qopp}           {Q_{\mathrm opp}}
\newcommand{\Lall}           {{\cal L}_i}
\newcommand{\Lsig}           {{\cal L}_i^{\mathrm signal}}

\newcommand{\pb}           {p_{\mathrm true}}
\newcommand{\mb}           {m_{\B}} % B mass, really means Bs
\newcommand{\Mi}           {M_i}

\newcommand{\li}           {l_i}
\newcommand{\lit}          {l_{{\mathrm true},i}}

\newcommand{\lengt}          {l_{\mathrm true}}
\newcommand{\sigLi}        {\sigma^l_i} % note similarity to \dl
\newcommand{\Bi}           {{\cal B}_i}
\newcommand{\tbgp}         {\tau_{\rm bg}^+}
\newcommand{\tbgn}         {\tau_{\rm bg}^-}
\newcommand{\ccbias}       {B^{\mathrm comb}_{\mathrm bias}}
\newcommand{\sysErr}       {\sigma^{\mathrm{systematic}}}
\newcommand{\sigt}         {\sigma^t}
\newcommand{\expt}         {t_{\mathrm {exp}}}
\newcommand {\MeVcc}      {{\rm{MeV}}}

\newcommand {\GeVc}       {{\rm{GeV}}}

\newcommand {\ps}         {\mbox{$\mathrm{\,ps}$}}
\newcommand {\psm}        {\ps^{-1}}

\newcommand {\mic}        {\mbox{$\mathrm{\,\mu m}$}}
\newcommand {\ts}         {\thinspace}
\newcommand {\bsym}       {\boldsymbol}
\newcommand {\mystrut}      {\rule{0pt}{3.2ex}}

\def \Items {\vspace {-.5ex} \begin{itemize}}
\def \endItems {\end{itemize} \vspace {-.5ex}}
\newcommand {\Caption}[1]  {\caption{\it #1}}
\newcommand {\addTable}[1] {\addtocontents{lot}{#1}}
\newcommand {\addFig}[1]   {\addtocontents{lof}{#1}}
\newcommand {\downto}
        {\mbox{ \begin{picture}(14,10)
                   \put(0,10){\line(0,-1){5.0}}
                   \put(2,5){\oval(4,4)[bl]}
                   \put(1,0){\makebox(0,0)[bl]{$\rightarrow$}}
                \end{picture} }}
\newcommand {\clnf}         {at the 95\% confidence level}
\begin{document}
% \pagenumbering{roman}
 
\begin{titlepage}
 
 \begin{center}
    {\large EUROPEAN ORGANIZATION FOR NUCLEAR RESEARCH}
 \end{center} 
 \begin{flushright}
   CERN-EP/2000-136 \\
   OPAL PR327 \\
   7 November 2000 \\
 \end{flushright}

  \begin{center}{\LARGE\bf
    A Study of $\bsym \Bs$ Meson Oscillation Using \\ 
    $\bsym \Ds$-Lepton Correlations\\}
  \end{center}
 
  \vspace{0.3in}
  \begin{center}
%     {\large\bf DRAFT 4}
  \end{center}

  \vspace{0.3in}
  \begin{center}
    {\Large\bf The OPAL Collaboration \large}
  \end{center}
 
\vspace{5 mm}

% \begin{center}{ \bf Authors: A. Harel, Y. Rozen, S. Tarem}\end{center}
% \begin{center}{ \bf Editorial board :
% R. Hawkings, C. Jones, J. McKenna, J. White}
% Richard Hawkings, Chris Jones, Janis McKenna, John White}
% \end{center}

  \vspace{0.1in}
  \begin{abstract}
    
From data collected around the $\Zzero$ resonance by the OPAL detector
at LEP, a sample of $\Bs$ decays was obtained using $\Dsl$
combinations, where the $\Ds$ was fully reconstructed in the
{\mbox{$\phipi$}}, {\mbox{$\KstK$}} and {\mbox{$\KKo$}} decay
channels or partially reconstructed in the {\mbox{$\philnux$}}
decay channel.
These events were used to study $\Bs$ oscillation.
The flavor (b or $\rm \bar{b}$) at decay was determined
from the lepton charge while the flavor at production
was determined
from a combination of techniques.
%four indicators: the momentum weighted jet
%charges for the candidate jet and the opposite jet and, where available,
%the charges of a prompt lepton in the opposite hemisphere and of a
%fragmentation kaon.
The expected sensitivity of the experiment is $\sensitivity\psm$.
The experiment was not able to resolve the oscillatory
behavior, and we deduced that the $\Bs$ oscillation frequency
$\dms > \limit\psm$ \clnf.
  \end{abstract}

  \medskip

  \vfill
  \begin{center}
    (Submitted to Euro. Phys. Jour.) \\
% {\large\bf Comments to aharel@physics.technion.ac.il by Friday Nov. 3rd}
  \end{center}
 
\end{titlepage}

%
%%%%%%%%%%%%%%%%%%%%%%%%%%%%%%%%%%%%%%%%%%%%%%%%%%%%%%%%%%%%%%%%%%%%%%%%%%%%%%%%
%
\begin{center}{\Large        The OPAL Collaboration
}\end{center}\bigskip
\begin{center}{
%begin authorlist PLEASE DO NOT DELETE THIS COMMENT
G.\thinspace Abbiendi$^{  2}$,
C.\thinspace Ainsley$^{  5}$,
P.F.\thinspace {\AA}kesson$^{  3}$,
G.\thinspace Alexander$^{ 22}$,
J.\thinspace Allison$^{ 16}$,
G.\thinspace Anagnostou$^{  1}$,
K.J.\thinspace Anderson$^{  9}$,
S.\thinspace Arcelli$^{ 17}$,
S.\thinspace Asai$^{ 23}$,
D.\thinspace Axen$^{ 27}$
% ...
%end authorlist PLEASE DO NOT DELETE THIS COMMENT
}\end{center}\bigskip
\bigskip
%begin institutes
$^{  1}$School of Physics and Astronomy, University of Birmingham,
Birmingham B15 2TT, UK
\newline
$^{  2}$Dipartimento di Fisica dell' Universit\`a di Bologna and INFN,
I-40126 Bologna, Italy
\newline
$^{  3}$Physikalisches Institut, Universit\"at Bonn,
D-53115 Bonn, Germany
\newline
% ...
$^{  5}$Cavendish Laboratory, Cambridge CB3 0HE, UK
\newline
% ...
$^{  9}$Enrico Fermi Institute and Department of Physics,
University of Chicago, Chicago IL 60637, USA
\newline
% ...
$^{ 16}$Department of Physics, Schuster Laboratory, The University,
Manchester M13 9PL, UK
\newline
$^{ 17}$Department of Physics, University of Maryland,
College Park, MD 20742, USA
\newline
% ...
$^{ 22}$Department of Physics and Astronomy, Tel Aviv University,
Tel Aviv 69978, Israel
\newline
$^{ 23}$International Centre for Elementary Particle Physics and
Department of Physics, University of Tokyo, Tokyo 113-0033, and
Kobe University, Kobe 657-8501, Japan
\newline
% ...
$^{ 27}$University of British Columbia, Department of Physics,
Vancouver BC V6T 1Z1, Canada
\newline
% ...
$^{  b}$ and Royal Society University Research Fellow
\newline
% ...
Tel Aviv 69978, Israel.
%end notes
\bigskip
%
%%%%%%%%%%%%%%%%%%%%%%%%%%%%%%%%%%%%%%%%%%%%%%%%%%%%%%%%%%%%%%%%%%%%%%%%

% \newpage
% \clearpage

% \pagenumbering{arabic}
%%%%%%%%%%%%%%%%%%%%%%%%%%%%%%%%%%%%%%%%%%%%%%%%%%%%%%%%%%%%%%%%%%%%%%%%
\section{Introduction}
\label{sec:intro}
%%%%%%%%%%%%%%%%%%%%%%%%%%%%%%%%%%%%%%%%%%%%%%%%%%%%%%%%%%%%%%%%%%%%%%%%
The phenomenon of ${\mathrm{B - \bar B}}$ mixing is well established.
In the case of the \Bd\ system,
the mass difference, $\dmd$, between the two 
mass eigenstates has been 
measured rather precisely~\cite{PDG}.
This mass difference gives the oscillation frequency
between $\Bd$ and $\mathrm{\overline{B}_d}$.
Although these measurements can be used to gain
information on the CKM matrix element $V_{\mathrm{td}}$,
this is hampered by large theoretical uncertainties
on both the meson decay constant, $f_{\mathrm B_d}$, and
the QCD bag model vacuum insertion parameter, $B_{\mathrm B_d}$~\cite{MJ2}.
This difficulty may be overcome if the \Bs\ oscillation
frequency, $\dms$, is also measured.
In this case, the CKM information can be extracted
via the relation:
\begin{equation}
\frac{\dms}{\dmd} = \frac{m_{\mathrm B_s}}{m_{\mathrm B_d}} \cdot 
 \frac{|V_{\mathrm{ts}}|^2}{|V_{\mathrm{td}}|^2}
   \cdot \frac{f^2_{\mathrm B_s} B_{\mathrm B_s}}{f^2_{\mathrm B_d} 
B_{\mathrm B_d}} \; ,
\end{equation}
where $m_{\mathrm B_s}$ and $m_{\mathrm B_d}$ are the \Bs\ and \Bd\ masses,
as the ratio of decay constants for \Bd\ and \Bs\ mesons is much
better known than the absolute values \cite{MJ2,MJ3}.
Information on $|V_{\mathrm{td}}|$ could then be extracted by 
inserting $|V_{\mathrm{ts}}|$, which is relatively 
well known~\cite{PDG}.

$\dms$ is predicted to be many times larger than $\dmd$ \cite{MJ2,MJ3}
and current lower limits support this theoretical predictions.
A large $\dms$ value leads to rapid oscillation thus presenting
experimental difficulties, which have prevented its measurement to date.
The most restrictive of the 
published limits~\cite{MJ4,MJ6,martin} 
indicates that $\dms > 9.6\psm$ \clnf~\cite{MJ6},
while the best limit from OPAL gives 
$\dms > 5.2\psm$ \clnf~\cite{martin}.

This paper describes an investigation of $\dms$ using a sample 
enriched in \Bs\
by reconstructing $\Dsl$ combinations\footnote{Throughout this paper charge
conjugate modes are implied.}. In OPAL this technique is expected to achieve a
sensitivity similar to that achieved by the inclusive
technique~\cite{martin}, since the better decay time resolution and
higher purity are offset by the lower statistics of an exclusive
analysis.

%%%%%%%%%%%%%%%%%%%%%%%%%%%%%%%%%%%%%%%%%%%%%%%%%%%%%%%%%%%%%%%%%%%%%%%%
\section{Analysis overview}
%%%%%%%%%%%%%%%%%%%%%%%%%%%%%%%%%%%%%%%%%%%%%%%%%%%%%%%%%%%%%%%%%%%%%%%%
\label{sec:overview}
The oscillation frequency of \Bs\ mesons was studied using exclusive
decays
of \Bs\ mesons into $\Dsl$ combinations. 
\Bs\ mesons were
reconstructed in the following four $\Ds$ decay channels as described 
in \cite{lifetimepaper}.
\vspace{-20pt}
\begin{tabbing}
  \hspace{2cm} \= \hspace{3.7cm} \= \hspace{2cm} 
  \= \hspace{2cm} \= \hspace{3cm} \\
  \> $\Bs \to \Ds\, \lp \, \nu$  \\
  \> $\phantom{\Bs \to \hspace{5pt}} \downto {\mathrm K^{\star 0}} \K^-$,
  \> ${\mathrm K^{\star 0}}\rightarrow \K^+\pi^-$ \> \\
  \> $\phantom{\Bs \to \hspace{5pt}} \downto \phi\, \pi^-$,
  \> $\phi\rightarrow \K^+\K^-$  \> \\
  \> $\phantom{\Bs \to \hspace{5pt}} \downto \Ko\, \K^-$,
  \> $\Ko\rightarrow \pi^+\pi^-$  \> 
  \> \\
  \> $\phantom{\Bs \to \hspace{5pt}} \downto \phi\lm\bar{\nu} \,\X $,
  \>$\phi\rightarrow \K^+\K^-$ \> 
  \>
\vspace{-5pt}
\end{tabbing}
The selection procedure of the event sample followed closely that of
\cite{lifetimepaper} and is briefly described in Section \ref{sec:selection} with an
emphasis on the changes made to suit the purpose of an
oscillation measurement. The background to the \Bs\ signal is described in
Section \ref{sec:bkg}. 

For each candidate we assigned a probability that it has mixed, i.e.,
its flavors ($\mathrm{b}$ or
$\bbar$) at production and at decay differ.
This probability was derived from the decay and
%For each event we assigned a probability that its decay and production
%flavors differ. This probability is derived from the decay and
production flavor tags, and we refer to it as a mixing tag 
(Section~\ref{sec:MT}).

In order to assign a likelihood of a candidate at a given $\dms$ value
we need, in addition to
the mixing tag, to reconstruct its decay time. Since the oscillation
measurement is highly sensitive to the decay time, we did not assume
a fixed Gaussian resolution on the decay time. We determined an
event-by-event probability
distribution for the decay time, which was derived from a Gaussian
probability distribution for the decay length
(Section~\ref{sec:dlest}), and from a non-Gaussian
probability distribution for the $\B$ candidate momentum 
(Section~\ref{sec:momest}).

In order to extract a lower limit on the oscillation frequency,
$\dms$, and to facilitate combination with other analyses we used the
amplitude fit method~\cite{moser}. This method
fits, for each value of $\dms$ checked,
a continuous parameter ${\cal A}$ which measures the size of the
component in the data oscillating at that particular value of $\dms$.
%compares the likelihood of
%the measurement arising from that particular value of $\dms$ and the
%likelihood of the measurement arising from a $\dms$ value too high to be
%measured. A $\dms$ value too high to be measured is easily simulated as an
%infinite one, resulting in infinite oscillations and no correlation
%between the production and decay flavors.
At the true $\dms$, the fitted value of
${\cal A}$ should be consistent with one,
while far below the true $\dms$, the expectation value for ${\cal A}$
is zero (see~\cite{gaelle} for additional details).
Therefore values of $\dms$ where ${\cal A}$ is below one and
inconsistent with one will be excluded.
The likelihood function and the fit results are described in Sections
\ref{sec:like} and \ref{sec:resfits}.
The systematic effects of the uncertainties on all the parameters used
in the amplitude fit were estimated by
repeating the amplitude fit with those parameters varied by one
sigma (Section \ref{sec:system}).
% performing a constrained fit assuming
%Gaussian errors on the input parameters (Section \ref{sec:system}).
Several checks of the method are described in Section~\ref{sec:checks}.
Finally our results, and the results of combining this measurement
with the previous OPAL measurement are summarized in
Section~\ref{sec:conclusion}.

%%%%%%%%%%%%%%%%%%%%%%%%%%%%%%%%%%%%%%%%%%%%%%%%%%%%%%%%%%%%%%%%%%%%%%%%
\section{Hadronic event selection and simulation}
%%%%%%%%%%%%%%%%%%%%%%%%%%%%%%%%%%%%%%%%%%%%%%%%%%%%%%%%%%%%%%%%%%%%%%%%
\label{sec:hadsel}

We used data collected by the OPAL detector~\cite{OPAL} at LEP between
1991 and 1995 running at center-of-mass energies in the vicinity of
the \Zzero\ peak with an operational silicon detector.
Hadronic \Zzero\ decays were selected using the number of tracks
and the visible energy in each event as in~\cite{evsel}. This
selection yielded 4.3 million hadronic events. 
In each event, tracks and electromagnetic clusters not
associated to a track were combined into jets, using the
JADE algorithm with the E0 recombination scheme.
Within this algorithm jets are defined by 
$y_{\mathrm {cut}}=0.04$~\cite{jade}.

%The primary vertex of the event was reconstructed using the 
%tracks in the event and the knowledge of the position and 
%spread of the $\ee$ collision point.
%
Monte Carlo samples of inclusive hadronic \Zzero\
decays and of the specific decay modes of interest were used to check
the selection procedure, mix tagging and fitting procedure. 
These simulated event samples included:
\begin{itemize}
% \Items
\item Samples of the four signal decay channels.
\item Hadronic $\Zzero$ decay samples, used to check the selection
  efficiencies and mix tagging of $\Zzero \rightarrow \qqbar$ decays,
  where q is a light quark (u, d, s or c).
%\Item $\Zzero \rightarrow \ccbar$ decay samples, used to check
%  the selection efficiencies and mix tagging of $\ccbar$ events.
\item $\Zzero \rightarrow \bbbar$ decay samples, used to check
  the selection efficiencies and the mixing tag of other background
  decays, such as partially reconstructed signal decays 
  (Section~\ref{sec:combMT}).
\item $\Zzero$ decay samples containing the specific decays 
  $\Bd \rightarrow \Dsp \D^-$, $\Bu \rightarrow \Dsp \D^0$, 
  $\Bd \rightarrow \Dsp \D^{-**}$,  $\Bu \rightarrow \Dsp \D^{0**}$,  
  $\Bu \rightarrow \Dsp \D^0 \pi$,  $\Bd \rightarrow \Ds \K^0 \lp \nu$
  and $\Bu \rightarrow \Ds \K^+ \lp \nu$, 
  used to check the selection efficiencies of $\Ds \lp$
  background decays (Section~\ref{sec:phys-bg}).
%\endItems
\end{itemize}

These samples were produced using the JETSET~7.4 parton shower Monte
Carlo generator~\cite{jetset} with the fragmentation function of
Peterson \etal ~\cite{peterson} for heavy quarks, and then passed
through the full OPAL detector simulation package~\cite{opalmc}.

%%%%%%%%%%%%%%%%%%%%%%%%%%%%%%%%%%%%%%%%%%%%%%%%%%%%%%%%%%%%%%%%%%%%%%%%
\section{Candidate selection}
%%%%%%%%%%%%%%%%%%%%%%%%%%%%%%%%%%%%%%%%%%%%%%%%%%%%%%%%%%%%%%%%%%%%%%%%
\label{sec:selection}

Three tracks were combined to form a $\Ds$ candidate and a lepton
(either $e$ or $\mu$) was added to form a $\Bs$ candidate. The four
tracks were required to be in the same jet.

The event selection and decay length reconstruction for this
analysis follow closely those of~\cite{lifetimepaper}.
Since the
decay time resolution is crucial in this analysis, an additional
requirement was made, demanding that the prompt lepton track (that is
the lepton directly from the $\B$ decay) had at least one 
associated hit in the silicon microvertex detector.
The photon conversion rejection has been updated to
use a neural network~\cite{gamcon}.
The event selection and reconstruction are
outlined briefly below:

Standard track quality cuts~\cite{opal-track-sel} were applied. 
Electrons were identified using a neural network~\cite{gccel}
% a proton rejection $\dEdx$ cut 
and a photon conversion rejection cut;
muons were identified
by associating central detector tracks with track segments in the muon
detectors and requiring a position match in two orthogonal
coordinates~\cite{gambblep}. For the other
reconstructed particles the probability
%$\dEdx$ (rate of energy loss due to ionisation) probability, $w_i$,
that the observed rate of energy loss due to ionisation ($\dEdx$) is 
consistent with the assumed particle hypothesis
was required to be greater than $1\%$. 

Additional channel dependent cuts included: momentum cuts, further
$\dEdx$ cuts,
invariant mass cuts on reconstructed intermediate particles, including
a loose cut on the invariant mass of the visible $\Bs$ decay products,
helicity angle cuts and a cut on the
angle between the $\Ds$ candidate and the prompt lepton candidate. 
See~\cite{lifetimepaper} for details.
 
In the $\KKo$ channel the mass of the two tracks forming the $\Ko$
candidate was constrained to the known $\Ko$ mass~\cite{PDG}.
Further constraints were applied to the $\Ds$ and
$\Ko$, in which the directions of the vectors between their production and
decay points were constrained to the reconstructed
momentum vectors. 
The lepton minimum momentum cut in this channel was $5\,\GeVc$.

Three vertices were reconstructed in the $x$-$y$ plane\footnote{The
  right-handed coordinate is defined such that the
  $z$-axis follows the electron beam direction and the $x$-$y$
plane is perpendicular to it with the $x$-axis lying approximately
horizontally. The polar angle $\theta$ is defined relative to the 
+$z$-axis, and the azimuthal angle $\phi$ is defined relative to the
+$x$-axis.}:
the $\ee$ interaction vertex, the $\Bs$ decay vertex
and the $\Ds$ decay vertex.  The $\ee$ interaction vertex was measured
using tracks
with a technique that follows any significant shifts in the $\ee$
interaction vertex position during a LEP fill~\cite{taulife}. 
The $\Ds$ decay vertex was fitted in the $r$-$\phi$ plane using all
the candidate tracks. The $\Bs$ decay vertex was formed by
extrapolating the candidate $\Ds$ momentum vector from its
decay vertex to the intersection with the lepton track.  

The $\Ds$ decay length is the distance between these two decay
vertices.  The $\Bs$ decay length was found by a fit between the
$\ee$ interaction vertex and the reconstructed $\Bs$ decay vertex using
the direction of the candidate $\Dsl$ momentum vector as a
constraint.  The two-dimensional projection of the $\Bs$ decay length
was converted into three dimensions using the polar angle that was
reconstructed from the momentum of the $\Dsl$.
Typical reconstructed decay length errors range from about 0.35\,mm
for the $\KKpi$, and $\philnu$ channels, to about twice this level
for the $\KKo$ channel.
% these numbers were checked in /bs/text/decayLengthErrorByChannel
In channels where the $\Ds$ was fully reconstructed the $\chi^2$ of
the $\Ds$ decay vertex
fit was required to be less than 10 (for one degree of freedom).
Finally, the reconstructed decay length error of the $\Bs$
candidate was required to be less than $0.2\,\mathrm{cm}$.
 
% % % % % % % % % % % % % % % % % % % % % % % % % % % % % % % % % % % % %
\subsection{ Results of $\bsym {\Ds \lp}$ selection}
% % % % % % % % % % % % % % % % % % % % % % % % % % % % % % % % % % % % %
 
%%%% MASS PLOT for Dsl%%%%
\begin{figure}[ptb]
  \begin{center}
  \begin{minipage}{0.9\textwidth}
     \epsfxsize=\textwidth
      \epsffile{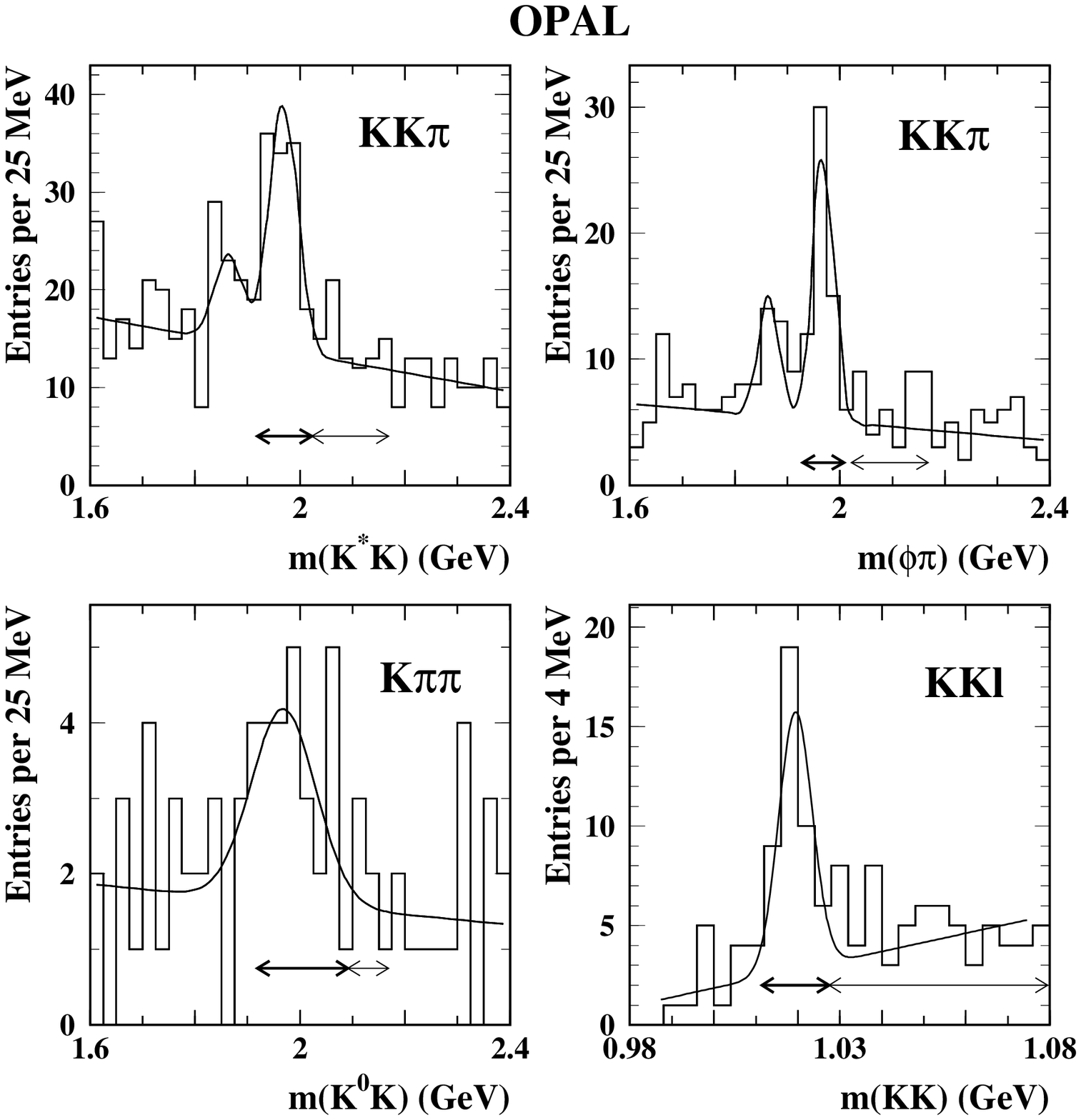}
  \end{minipage}
    \vspace{-0.2in} 
    \addFig{Invariant mass distributions}
    \Caption{Invariant mass distributions from the 
      four $\Ds$ decay channels.  In each plot,
      the result of the fit to the signal and possible 
      satellite peaks, as described in the text, is overlaid as 
      a solid line. The bold arrow to the left shows the signal region
%      ranges used in the oscillation fit and 
      and the lighter arrow to the right shows the sideband region. }
     \label{fig:Dsmass}
  \end{center}
\end{figure}

The invariant mass distribution obtained in each of the $\Ds$ decay
channels is shown in Figure~\ref{fig:Dsmass}.  
Each invariant mass distribution was fitted to a Gaussian
distribution describing the signal and a linear parameterization for
the combinatorial background. The mean of the Gaussian distribution
was fixed to the
nominal $\Ds$ mass, $1968.5\,\MeVcc$~\cite{PDG}, for the hadronic
channels and to the nominal $\phi$ mass,
$1019.413\,\MeVcc$~\cite{PDG}, for the semileptonic channel.
In the $\KKp$ distributions, a second Gaussian distribution was used
to parameterize contributions
from the Cabibbo suppressed decay $\D^-\to\KKpi$.  The mean of the second
Gaussian distribution was fixed to the nominal $\D^-$ mass,
$1869.3\,\MeVcc$~\cite{PDG}, and the width was constrained to be the
same as that of the $\Ds$ peak. 
The combinatorial background in the semileptonic channel was refitted
 to account for the kinematical threshold as in~\cite{btophi}. The
 choice of the background parameterization was found to have a
 negligible effect on the fitted amplitude.
%old (still true for phi-pi oddly enough) By integrating the tail of the peak
%due to the $\D^-\to\KKpi$ decays in this $\Ds$ signal region, the
%contamination from this source is found to be negligible.
For each channel, the fitted width was consistent with the
expected detector resolution. 
The contamination due to the $\D^-\to\KKpi$ decays was estimated from
simulated events as explained in Section~\ref{sec:combMT}.
The results of these fits are summarized in Table~\ref{tab:signal}.

\begin{table}[bth]
  \vspace{0.2cm}
  \centering
  \begin{tabular}{|l|c|c|c|c|c|}
    \hline
      Decay   &  Candi- & Comb.          & Signal       &
      Sideband        & Estimated \\
      channel &  dates& fraction         & region (\MeVcc)& 
      region (\MeVcc) & signal    \\
    \hline \hline
   $\K^*\K$   & 125         &  0.455$\pm$0.026 & 1918.5 -- 2022.1 &
      2022.1 -- 2168.5 & 53.8$\pm$5.0\\
   $\phi\pi$  & 54          &  0.277$\pm$0.027 & 1929.5 -- 2007.5 & 
      2022.1 -- 2168.5 & 30.9$\pm$2.7\\
   $\Ko\K$     & 24          & 0.467$\pm$0.092 & 1918.5 -- 2091.3 & 
      2091.3 -- 2168.5 & 10.1$\pm$2.3\\
   $\phi\ell$ & 41          &  0.243$\pm$0.039 & 1011.4 -- 1027.4 &
      1027.4 -- 1079.4 & 21.4$\pm$2.8\\
    \hline
   Total      & \ncanddsl   &  0.386$\pm$0.018 &  &  & \ncandbs$\pm$\ncandbserr \\
    \hline
  \end{tabular}
  \addTable{Mass fit and event selection results}
  \Caption{
    Results of the mass fits to all the signal
    channels. The second column shows the
    number of events selected in the signal region as defined in the
    text. The fitted fraction of combinatorial background for
    events selected in the signal region is given in the third column,
    together with the associated statistical error. The fourth column
    gives the mass range for the signal region, which corresponds to
    about twice the fitted width around the nominal $\Ds$ mass for the
    hadronic channels, and to about twice the width around the nominal
    $\phi$ mass for the semileptonic channel. The fifth
    column gives the mass range for the sideband region.
    The last column gives the estimated number of correctly
    reconstructed $\Bs \rightarrow \Dsl$ signal events in the signal
    region, where both combinatorial and other background discussed
    in~\ref{sec:bkg} were subtracted. The errors are from the
    uncertainties in the combinatorial and non-combinatorial
    background subtraction.}
    \label{tab:signal}
    \vspace{0.3cm}
\end{table}

No significant peaks were observed in the mass distributions for
same-sign $\Dslwr$ combinations in the fully reconstructed
decay channels \Kstar\Km, $\phipi$ and $\KKo$.  
% yep, checked that out 30/9/99 see massPlotsWrongSign

The signal and sideband regions are defined in Table~\ref{tab:signal}
and shown in Figure~\ref{fig:Dsmass}.
The $\Dsl$ combinations selected for the oscillation fit were 
from the signal regions. $\ncanddsl$
such $\Dsl$ candidates were observed.
The candidates selected in the mass sideband regions were used to
estimate the lifetime characteristics of the combinatorial background.
$\nsideband$ such sideband candidates were observed. 
The regions were chosen to represent the decay time distribution of
background under the signal as in~\cite{lifetimepaper}.

% % % % % % % % % % % % % % % % % % % % % % % % % % % % % % % % % % % % %
\subsection{Background to the $\bsym {\Bs \to \Ds \lp}$ signal}
% % % % % % % % % % % % % % % % % % % % % % % % % % % % % % % % % % % % %
\label{sec:bkg}
 
Potential sources of non-combinatorial background to the \Bs\ signal
considered here include decays of other B~hadrons that can yield a
$\Dsl$ final state or other final states that were
misidentified as a $\Ds$ meson.  Other sources are a $\Ds$
 combined with a hadron that has been misidentified
as a lepton, and random associations of a $\Ds$ with
a genuine lepton. Finally, there is combinatorial background from
misreconstructed $\Ds$ mesons in the hadronic channels and
misreconstructed $\phi$ mesons in the semileptonic channel.
The various background sources, and the calculation of their
contributions relative to that of the signal, are discussed below.
 
%       %       %        %       %       %       %       %       %       %
\subsubsection {$\bsym {\Ds \lp}$ Background}
%       %       %        %       %       %       %       %       %       %
\label{sec:phys-bg}

The event sample includes properly reconstructed
$\Dsl$ combinations that do not arise from $\Bs$ decay. Two decay
modes of $\Bd$ and $\Bu$ mesons were considered: 
%\Items
\begin{itemize}
  \item[(a)] $\Budbar \to \Ds \D \X$, 
             \mbox{$\D\to\lp\nu\X$} 
             (where $\D$ is any non-strange charm meson).
  \item[(b)] \mbox{$\Bud\to \Ds \K \lp\nu\X$}, 
             where $\K$ also represents excited kaons.
%\endItems
\end{itemize}
Note that in decay mode (a) a negative lepton would indicate a $\Bu$
 or $\Bd$ meson, whereas in signal decay and in decay mode (b) a
 negative lepton would indicated a $\Bsbar$, $\Bubar$ or $\Bdbar$.
% Decay mode (a) of $\Bu$ and $\Bd$ mesons produces a negative lepton,
% whereas the $\Bs$ signal decay and decay mode (b) of $\Bu$ and $\Bd$
% mesons produce a positive lepton.

Mode (a) includes two body decays $\Budbar\to\Ds\D$ for which a branching
ratio measurement exists, and three body decays $\Budbar\to\Ds\D\Xnop$ for
which no branching ratio measurement exists. The measurement
$\Br(\Budbar\to\Ds\Xnop) = 0.100 \pm 0.025$~\cite{PDG} provides an upper
limit on the sum of these modes. To estimate this mode's contribution 
to the $\Ds \lp$ sample, the semileptonic branching ratios for the 
different non-strange D hadrons are weighted according to their
abundance in $\Budbar\to\Ds\D$ decays. 
The efficiency for this analysis to select $\Ds \lp$ combinations 
from these modes was taken from Monte Carlo and included in the
calculation.  The possibile contribution from analogous decay modes of
b-baryons into $\Ds^{(*)} \Lc$ 
was included in the calculation although none have been observed to
date. These modes account for $0.143 \pm 0.050$ of the selected $\Ds
\lp$ combinations, and $0.619 \pm 0.045$ of this background comes from
$\Bd$ decays. 

%
%Through these modes $\Bd$ decays account for $0.089 \pm 0.036$
%of the selected $\Ds \lp$ combinations, and $\Bu$ and b-baryon decays
%account for a further $0.055 \pm 0.023$.

Mode (b), $\Bud\to\Ds\K\lp\nu\X$, has not been observed and only an
upper limit 
of $0.009$ (90\%CL)~\cite{PDG} exists. However, a branching ratio for this 
mode can be calculated from the fraction of 'lower vertex' ($\Ds$ not
produced from the virtual W)
$\B\to\Ds\Xnop$~\cite{PDG} combined with the inclusive B
semileptonic decay fraction. 
This yields $\Br(\B\to\Ds\K\lp\nu\X) = 0.0018\pm0.0009$ which is 
consistent with the above upper limit as well as with the theoretical upper 
limit~\cite{Dsklnu-lim}.
Monte Carlo events were used to determine the
selection efficiencies for these background modes relative to that of the
signal mode. Analogous baryonic modes were included in this calculation as 
well. These modes account for $0.065 \pm 0.035$ of the selected $\Ds
\lp$ combinations, and $0.470 \pm 0.038$ of this background comes from
$\Bd$ decays. 
%
% Through these modes $\Bd$ decays account for $0.031 \pm 0.018$
% of the selected $\Ds \lp$ combinations, and $\Bu$ and b-baryon decays
% account for a further $0.034 \pm 0.020$.
% of the selected $\Ds \lp$ combinations.

% Due to the dependance of these estimates on the $\Bs$ production fraction
% $f_s$

%       %       %        %       %       %       %       %       %       %
\subsubsection {Other background}
%       %       %        %       %       %       %       %       %       %
\label{sec:otherbg}
The background from genuine $\Ds$ particles that
% (that preferable to which, though both legit.)*2 since this are
% restrictive clauses.
were combined with a hadron that was misidentified as a lepton can be
estimated from the invariant mass spectrum of 
combinations of same-sign charm candidate and lepton candidate pairs.
Assuming that misidentified hadrons are equally likely in both
charges, the same number of $\Ds$+fake lepton should exist with the
correct charge correlation.
%This assumes that random
%combinations are equally likely to have right and wrong charge
%correlations. 
For each channel in which the charm hadron is fully
reconstructed, no significant excess of same sign $\Ds$ signal exists.
This is in agreement with what has been found in a related analysis that
has greater statistical significance~\cite{opalBzeroBplus}.
This background source was therefore neglected.

In the channel $\Ds\to\phi\lp\nu\X$, where the charm hadron was partially 
reconstructed from a semileptonic decay channel, there was additional 
background to consider. 
%As noted above, the sample selection for this analysis follows closely
%that of reference
This background includes the accidental combination of
a $\phi$, produced in fragmentation, with two leptons
that arise from either $\B \to \D \ell \Xnop, \D \to \ell \Xnop$ or 
$\B \to \rm J/\psi$ decays and candidates from hadrons misidentified
as leptons.
In~\cite{lifetimepaper} it was estimated that
the fraction of candidates in the $\phi$ signal region that arise
from this background
%out of $37$ events found in the $\phi$ signal region,
% $5\pm2.1$ candidate arise from background
particular to the $\Ds\to\phi\lp\nu\X$ channel is 
$\fother = 0.135 \pm 0.057$, an estimate used here as well.
%\begin{enumerate}
%\item $2.5\pm0.5$ candidates from the accidental combination of
%a $\phi$, produced in fragmentation, with two leptons
%which arise from a semileptonic bottom hadron decay, followed by a
%semileptonic charm hadron decay. 
%\item $0.5\pm0.3$ candidates from leptons arising from $\rm J/\psi$
%  decays,  which are then combined with a $\phi$ (either
%from fragmentation or from a b~hadron decay).
%\item $2\pm2$ candidates from hadrons misidentified as leptons.
%\end{enumerate}

The non-combinatorial background sources mentioned above were
expected to contribute a total of $\nbsphysBG$ events to the $\Dsl$ signal.  
The background subtracted number of $\Dsl$ signal candidates was 
therefore 
%which proceed through the chosen decay chains is then
$ N({\Bs\to\Dslnux})\ = \ \ncandbs \pm \ncandbserr $, as given in
Table~\ref{tab:signal}.

%%%%%%%%%%%%%%%%%%%%%%%%%%%%%%%%%%%%%%%%%%%%%%%%%%%%%%%%%%%%%%%%%%%%%%%%
\section{Proper decay time reconstruction}
%%%%%%%%%%%%%%%%%%%%%%%%%%%%%%%%%%%%%%%%%%%%%%%%%%%%%%%%%%%%%%%%%%%%%%%%
\label{sec:time}

% The $\Bs$ lifetime is related to the
% observed decay lengths.
The true $\Bs$ proper decay time, $t$, is derived
% from the candidate's decay length and momentum 
using the relation:
\begin{equation}
t = \frac {\lengt \cdot \mb } {\pb}~,
\label{equ:propertime}
\end{equation}
where $\lengt$,\ $\pb$ and $\mb$ are the $\Bs$
candidate's true decay length, true momentum and
nominal mass respectively. From the measured decay length we derive a
Gaussian probability distribution for $\lengt$, as described in
Section~\ref{sec:dlest}. We also derive a non-Gaussian probability
distribution for $\pb$, as described in Section~\ref{sec:momest}.

%       %       %        %       %       %       %       %       %       %
\subsection{Decay length estimation}
%       %       %        %       %       %       %       %       %       %
\label{sec:dlest}

\begin{figure}[tb]
\centering
\epsfxsize=11cm
\begin{center}
    \leavevmode
    \epsffile{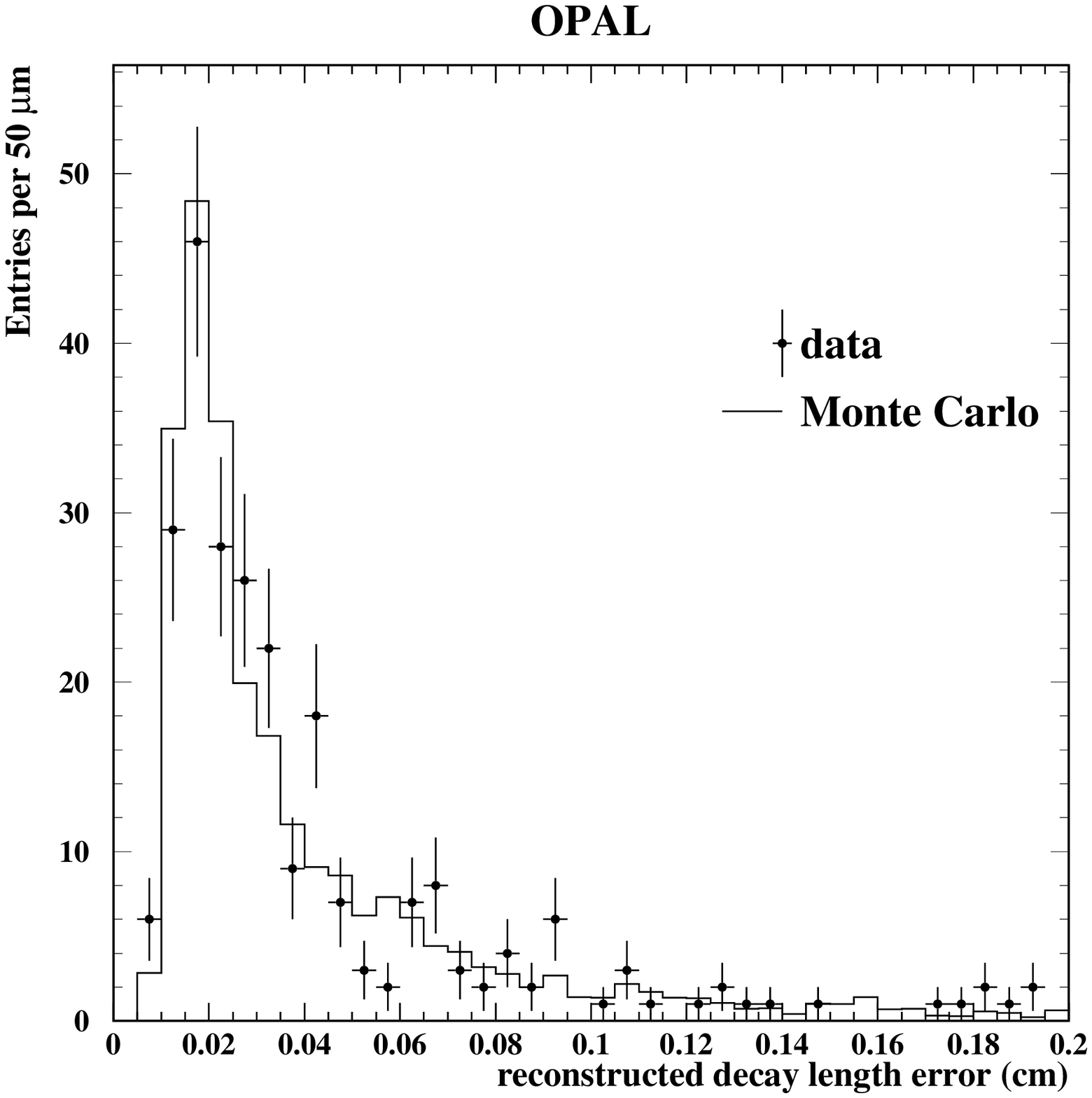}
\end{center}
\vspace{-5 mm}
\addFig{Decay length error reconstruction in data and simulation}
\Caption{ 
Distribution of reconstructed decay length errors for
selected events. The points with the error bars represent the data
while the histogram represents the simulated events. 
% the statistical errors on MC are typically about half than on data.
% the MC is weighted 4 signal channels & 1 type of comb. bkg.
}
\label{fig:deb}
\end{figure}

The $\Bs$ candidate's decay length is reconstructed as described
in Section~\ref{sec:selection}.
%in reference~\cite{lifetimepaper}. 
Using simulated events it was found that the decay length
reconstruction is biased and that
the decay length errors reconstructed by this method were overly
optimistic by a factor of about 1.4.
On average the reconstructed decay length was bigger than the true
decay length by $24 \pm 12~\mic$. We corrected for this bias, which
is 6\% of the average decay length resolution, and less than 1\% of
the average decay length.
The distribution of the reconstructed decay length errors in the data
and in simulated events is similar, as shown in Figure~\ref{fig:deb}. 
Using simulated signal events we fitted the ratio between
the correct decay length error, $\dl$, and the reconstructed
decay length error, $\dlrec$, as a linear function
of $\dlrec$. 
Simulated signal events from all four decay channels were used,
and the dependance of $\dl$ on $\dlrec$ was similar in all signal channels.
% see ~/bs/text/dlErr.sources
We used this function to correct the decay length error in the
likelihood function calculation, and used the fitted uncertainty on
this function as a systematic error.
% but was found to be small relative to the systematic
% uncertainties for this fit which are due to uncertainties in the
% resolution of the tracking detectors (Section~\ref{sec:system}).

%

%
% \begin{figure}[tb]
%  \centering
%  \epsfxsize=14cm
%  \begin{center}
%    \leavevmode
%    \epsffile{../text/dlerr.factor.eps}
%  \end{center}
%  \vspace{-5 mm}
%  \addFig{decay length resolution ratio}
%  \caption{ 
%    The fitted resolution ratio (as defined in the text) as a
%    function of the bin's average reconstructed decay length
%    resolution. The result of the fit is shown by the solid line,
%    while one sigma errors on both sides of it are shown as dashed
%    lines.
%    }
%  \label{fig:widfit}
% \end{figure}

%       %       %        %       %       %       %       %       %       %
\subsection{Momentum estimation}
%       %       %        %       %       %       %       %       %       %
\label{sec:momest}
Since the prompt neutrino produced in the $\Bs$ candidate's decay,
and in some cases additional decay products, are not reconstructed,
there is no direct measurement of the candidate's true momentum $\pb$.
The binned probability distribution of the candidate's true momentum,
${\cal B}(\pb)$, is estimated on an event-by-event basis using a
probability distribution based on the reconstructed $\Bs$ candidate
(${\cal B}_1$, Section~\ref{sec:mom1}) and a
probability distribution based on the recoil to the candidate,
i.e. the other tracks and clusters in the event (${\cal B}_2$,
Section~\ref{sec:mom2}). The two probability
distributions were then used to calculate ${\cal B}$ using:
\begin{equation}
{\cal B}(\pb) =
\frac{{\cal B}_1(\pb) \cdot {\cal B}_2(\pb)}
{\sum_{i=1}^n {\cal B}_1(p_i) \cdot {\cal B}_2(p_i) }~,
\label{equ:momJoin}
\end{equation}
where $ n $ is the number of momentum bins.

% {\Sigma_{i=1}^{nBin} {\cal B}_1(P_i) \cdot {\cal B}_2(P_i) } $$

%       %       %        %       %       %       %       %       %       %
\subsubsection{Candidate based momentum distribution (${\cal B}_1$)}
%       %       %        %       %       %       %       %       %       %
\label{sec:mom1}
We calculate a probability distribution for the $\Bs$ candidate's
true energy, $E_{\mathrm B}$, using the reconstructed invariant 
mass, $m_{{\mathrm D}\ell}$, and energy, $E_{{\mathrm D}\ell}$, of the $\Ds$
lepton combination as experimental inputs, following the
method presented in~\cite{opalBzeroBplus}.

A Bayesian approach is used for which an $a \, priori$ knowledge of the
$\Bs$ candidate's energy spectrum is required. This $a \, priori$
spectrum, $P(E_{\mathrm B})$, was derived from Monte Carlo.
% The method is not sensitive to details of the simulation.
Applying two body decay kinematics, the observable energy,
$E_{{\mathrm D}\ell}$, is given in the laboratory frame by:
\begin{equation}
  E_{{\mathrm D}\ell} =  \frac{\gamma_{\mathrm B}}{2m_{\mathrm B}}
    ( \Sigma + \beta_{\mathrm B} \Delta \cos \theta^{*}_{\B} )~,
\end{equation}
where $\theta^{*}_{\B}$ is the angle between the 
${\mathrm D}\ell$ flight direction and the boost
vector in the $\Bs$ candidate's rest frame, $\Sigma = m^2_{\mathrm B}
+ m^2_{{\mathrm D}\ell}$~, $\Delta = m^2_{\mathrm B} - m^2_{{\mathrm
    D}\ell}$~, $\beta_{\mathrm B}$ and $\gamma_{\mathrm B}$ are the
boost parameters of the $\Bs$ candidate in the laboratory frame. 

The distribution in \mbox{$\cos \theta^{*}_{\B}$} is uniform (because 
the B meson is a pseudoscalar particle),
therefore $E_{{\mathrm D}\ell}$ is distributed uniformly 
between $\frac{\gamma_{\mathrm B}}{2m_{\mathrm B}} 
( \Sigma - \beta_{\mathrm B} \Delta)$ and 
$\frac{\gamma_{\mathrm B}}{2m_{\mathrm B}} 
( \Sigma + \beta_{\mathrm B} \Delta)$.

We then used the fact that $E_{\mathrm B}$ is independent of the $\D\ell$
invariant mass to get 
$P(E_{\mathrm B},m_{{\mathrm D}\ell}) = P(E_{\mathrm B}) \cdot
P(m_{{\mathrm D}\ell})$, together with Bayes theorem to obtain the
formula:
\begin{equation}
 P(E_{\mathrm B} | E_{{\mathrm D}\ell}, m_{{\mathrm D}\ell} ) = 
\frac{P(E_{{\mathrm D}\ell} | E_{\mathrm B},m_{{\mathrm D}\ell})
  P(E_{\mathrm B},m_{{\mathrm D}\ell})}{\int P(E_{{\mathrm D}\ell} |
  E^{'}_{\mathrm B},m_{{\mathrm D}\ell})
  P(E^{'}_{\mathrm B},m_{{\mathrm D}\ell}) \ud E^{'}_{\mathrm B}} =
\frac{P(E_{{\mathrm D}\ell} | E_{\mathrm B},m_{{\mathrm D}\ell})
  P(E_{\mathrm B})}{\int P(E_{{\mathrm D}\ell} | E^{'}_{\mathrm
  B},m_{{\mathrm D}\ell})
  P(E^{'}_{\mathrm B}) \ud E^{'}_{\mathrm B}}~.
\end{equation}

The momentum probability density, ${\cal B}_1$, is then derived from
the energy probability density.
% this Is slightly unfair: the real error formula is in energy space
% (not momentum) and so the error numbers are really on energy!
Using simulated signal decays, it was found that
on average the expectation value of ${\cal B}_1$ was smaller than the true
momentum by $0.24 \pm 0.06~\GeVc$. We corrected for this bias, which
is less than 1\% of the average candidate momentum.

%       %       %        %       %       %       %       %       %       %
\subsubsection{Recoil based momentum distribution (${\cal B}_2$)}
%       %       %        %       %       %       %       %       %       %
\label{sec:mom2}
Another way of obtaining a good estimate of the $\Bs$ candidate's
momentum is to use our knowledge of the total center of mass energy
$\Ecm$, which is twice the LEP beam energy. 
We calculate the $\Bs$ candidate's energy using this constraint
and the recoil mass of the rest of the event, using the relation:
\begin{equation}
E_{\mathrm B} =  \frac{\Ecm^2 + m_{\mathrm B}^2 - m_{\mathrm rec}^2}
{2 \Ecm}~,
\end{equation}
where $m_{\mathrm rec}$ is the recoil mass calculated using all
tracks and unassociated electromagnetic clusters in the event
$except$ the reconstructed $\Bs$ decay products, and $m_{\mathrm B}$
is the nominal $\Bs$ mass.
In this calculation all tracks were assigned the pion mass
and all neutral clusters were taken as massless.
The candidate's momentum is then given by $\sqrt{E_{\mathrm B}^2 -
  m_{\mathrm B}^2}$.

\begin{figure}[tb]
\centering
\epsfxsize=11cm
\begin{center}
    \leavevmode
    \epsffile{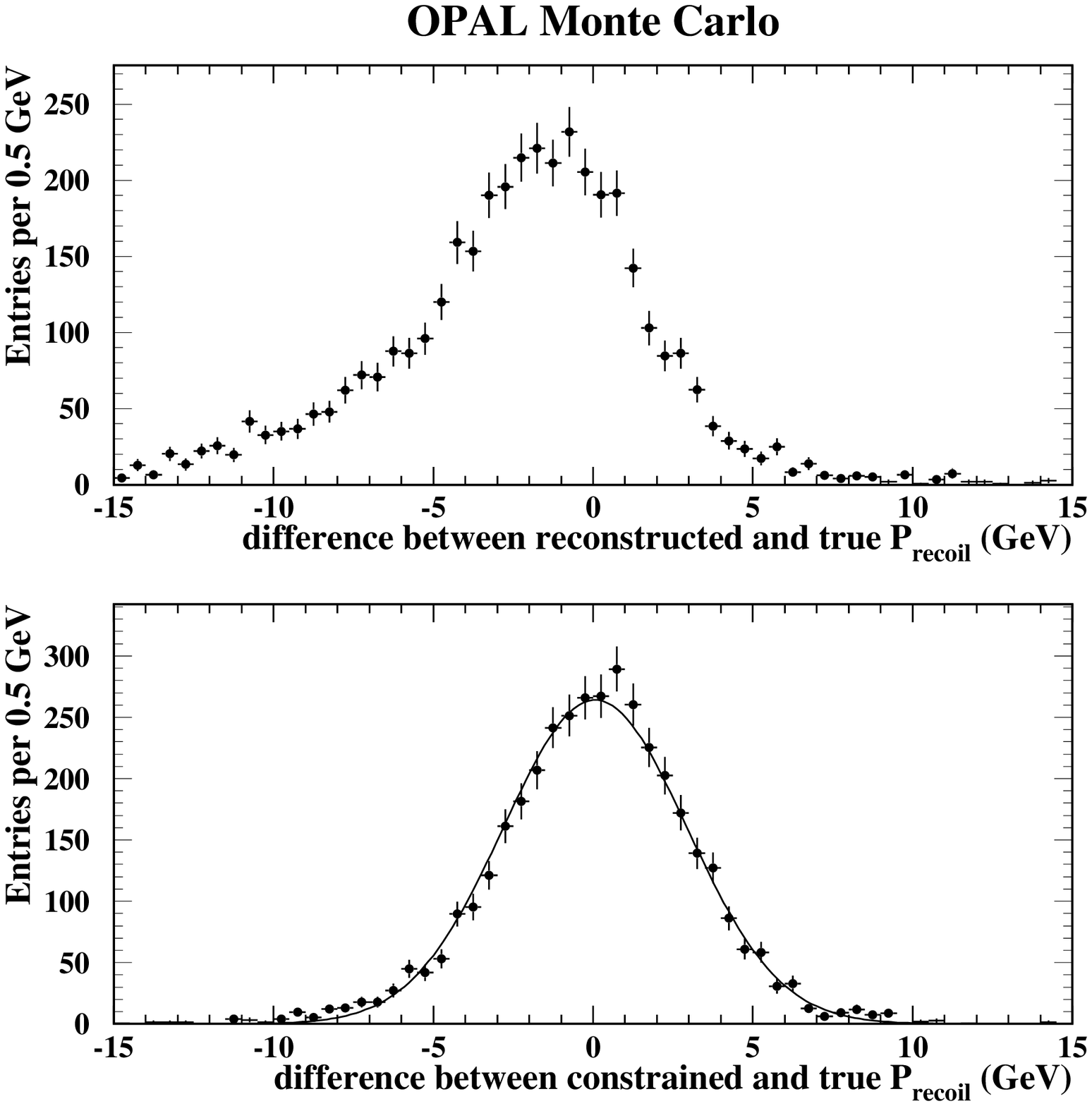}
\end{center}
\vspace{-5 mm}
\addFig{Constraining the recoil momentum estimate}
\Caption{ 
Comparison of errors in the momentum estimate ${\cal B}_2$ before
(upper) and after (lower) the
visible energy correction of $m_{\mathrm rec}$ on simulated signal events.
The points with the error bars represent the simulated events
while the solid line shows a Gaussian fit to the points.  
}
\label{fig:recmom}
\end{figure}

Monte Carlo studies have shown that the accuracy of this estimate can
be improved by rescaling $m_{\mathrm rec}$ according to
the visible energy, calculated using all
tracks and unassociated electromagnetic clusters in the event, and
$\Ecm$ (see Figure~\ref{fig:recmom}).
% We apply a correction to $m_{\mathrm rec}$ to reflect this constraint.
% that is afinite linear in $E_{vis}-\Ecm}$.

Figure~\ref{fig:recmom} shows that the difference between the corrected
reconstructed momentum and the true simulated momentum ($\pb -
\pbrec$) is well described by a Gaussian distribution whose center is
at $0.31 \pm 0.05~\GeVc$. We corrected for this bias, which is less
than 1\% of the average candidate momentum.
Therefore ${\cal B}_2$ was chosen as a Gaussian distribution around the
reconstructed momentum minus the bias ($\pbrec - 0.31~\GeVc$) with a
width of 2.88 $\GeVc$,
i.e. the fitted Gaussian width of $\pb - \pbrec$ in simulated signal
events.
% see ~/bs/text/momentumNumbers and ~/bs/text/timeBias

%
%

%       %       %        %       %       %       %       %       %       %
\subsubsection{Results of momentum estimation}
%       %       %        %       %       %       %       %       %       %
\label{sec:momestres}
The width (RMS) of the ${\cal B}_1$ distribution varies greatly
between events, the average width on simulated signal events is 3.7
$\GeVc$. As stated above the ${\cal B}_2$ distribution has a single
width of 2.88 $\GeVc$ for all events. The average width of the
combined distribution ${\cal B}$ on simulated signal events is 2.30
$\GeVc$.

The small individual biases on ${\cal B}_1$ and ${\cal B}_2$ were
corrected before combining them according to
Equation~\ref{equ:momJoin}.
% since we dont fix this directly (but rather through time bias), why
% mention it?
% Using simulated signal events it was found
%that the resulting combined distribution has a
%residual bias of $0.29 \pm 0.05~\GeVc$, which is less than 1\% of the
%average candidate momentum. 
It was verified on the simulated signal
events that after correcting for both biases, ${\cal B}$ is indeed
a reasonable representation of the true
probability distribution for the candidate's momentum.

%       %       %        %       %       %       %       %       %       %
\subsection{Results of proper decay time estimation}
\begin{figure}[tb]
\centering
\epsfxsize=11cm
\begin{center}
    \leavevmode
    \epsffile{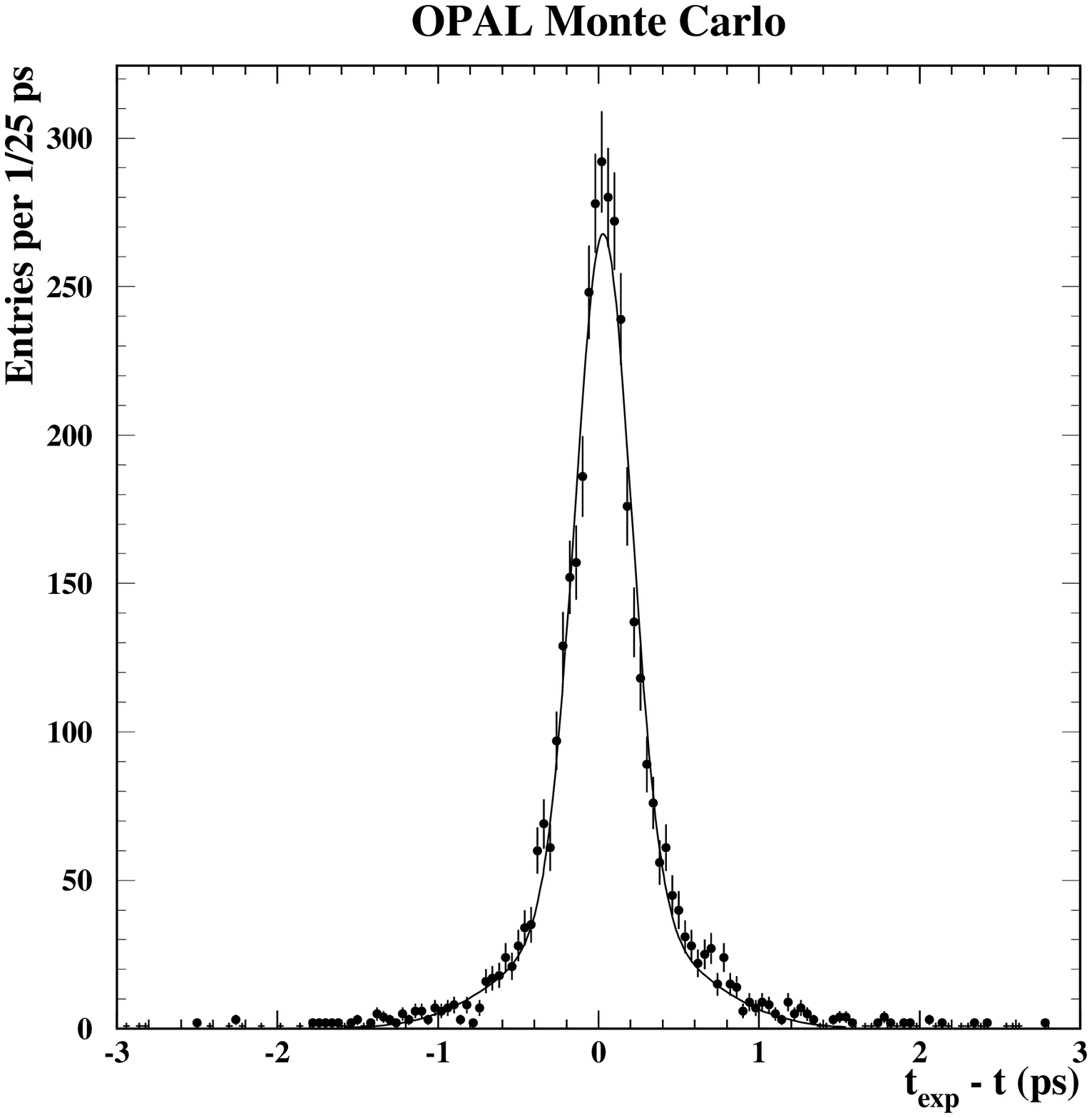}
\end{center}
\vspace{-5 mm}
\addFig{Decay time resolution}
\Caption{ 
Difference between the expectation value of the estimated probability
distribution of the true proper decay time, $\expt$, and the true
proper decay time, $t$.
The points with the error bars represent the simulated events
while the solid line shows a double Gaussian fit to the points.
68\% of the events lie within the narrow Gaussian distribution whose
width is $0.175\ps$. The width of the wide Gaussian distribution is
$0.52\ps$.  
}
\label{fig:dtexp}
\end{figure}

%       %       %        %       %       %       %       %       %       %
\label{sec:dtres}
The distribution of the true proper decay time is estimated by
combining the decay length estimate 
(\ref{sec:dlest}) and the momentum estimation (\ref{sec:momest})
according to Equation~\ref{equ:propertime}, after correcting for their
small biases.
Figure~\ref{fig:dtexp} shows that for $68\% \pm 5\%$ of simulated
signal events the difference between the expectation value of
the reconstructed true proper decay time distribution, $\expt$, 
and the true decay time, $t$, is well characterized by a Gaussian
distribution of width $0.175 \pm 0.011\ps$.
% A comparison of the expectation value of
% the reconstructed true proper decay time distribution, $\expt$, 
% and the true decay time, $t$, plotted in
% Figure~\ref{fig:dtexp} shows that for $68\% \pm 5\%$ of simulated
% signal events the difference is well characterized by a Gaussian
% distribution of width $0.175 \pm 0.011\ps$.
This fit is shown for information only, and was not used
in the oscillation fit likelihood.

We classified simulated signal events according to the RMS of their
reconstructed true proper decay time distribution, $\sigt$. We found
that in events with low $\sigt$ the expectation value of
the reconstructed true proper decay time distribution, $\expt$, tends
to be smaller than the true proper decay time, $t$. In 
events with high $\sigt$, $\expt$ tends to be bigger than $t$. 
This residual bias is about 2\% of the average proper decay time.
We fitted the proper decay time reconstruction bias as a linear
function of $\sigt$ 
($t_{\rm bias}^{\rm slope} \dot \sigt + t_{\rm  bias}^0$),
and treated the fitted uncertainties as sources of systematic
uncertainty.

%%%%%%%%%%%%%%%%%%%%%%%%%%%%%%%%%%%%%%%%%%%%%%%%%%%%%%%%%%%%%%%%%%%%%%%%
\section{Mixing tag}
%%%%%%%%%%%%%%%%%%%%%%%%%%%%%%%%%%%%%%%%%%%%%%%%%%%%%%%%%%%%%%%%%%%%%%%%
\label{sec:MT}

%To distinguish mixed from unmixed $\Bs$ decays, we need to determine the 
%b~flavor of the $\Bs$ both at production and at decay. 
The mixed and unmixed $\Bs$ decays were distinguished by determining
the b~flavor of the $\Bs$ (whether it contains a b or
$\overline{\mathrm{b}}$ quark) both at production and at decay.
The b~flavor at decay was inferred from the charge of the prompt
lepton in the $\Dsl$ combination. The initial
b~flavor was tagged by a combination of the charge of a lepton in the
hemisphere opposite the $\Bs$ 
candidate, the charge of a fragmentation kaon in the candidate
hemisphere, and jet charge measures from both the
candidate hemisphere and the opposite hemisphere. The available tags
in each hemisphere were combined into a 
measure of the probability that the candidate was produced as a $\Bs$
and then 
the probabilities from the two hemispheres were combined into a single
probability.
% measure by neglecting any correlations between the tags
% above those
% arising from having both tags tag the same quantity. On simulated
% events these neglected correlations were found to be consistent with
% zero and of the order of 1\%.
% actually by conditioning by flavor remove the known correl.
% and the 2 Cor (=Cov/(V1*V2)^-.5) found were .0136 & -.0053 
% giving Cor=.004@.009. No computer file, it's an old piece of paper.
It was verified that any unwanted correlations between the flavor
tags of the two hemispheres were negligible.
The mixing probability is derived from the production flavor
probability and the decay flavor.

%       %       %        %       %       %       %       %       %       %
\subsection{The $\bsym \Bs$ candidate hemisphere}
%       %       %        %       %       %       %       %       %       %
\label{sec:MTcand}

The $\Bs$ production flavor was measured in the $\Bs$ candidate's
hemisphere by the jet charge and, 
%Up to two measures of the produced b~flavor of the $\Bs$ candidate were
%identified in 
%the hemisphere containing the $\Bs$ candidate: the jet charge and,
where available, the charge of a kaon from the fragmentation process.

The jet charge of the jet containing the $\Bs$ candidate was calculated as
\begin{equation}
\Qsame = {1 \over {(E_{\rm beam})^{\kappa}}} \cdot \sum_{{\it
    i}=1}^{\it n} {\it q}_{\it i}
\cdot {\it (p_i^l)^{\kappa}}~,
\end{equation}
where $p_i^l$ is the longitudinal component of the momentum of
particle $i$ with respect to the jet axis, $q_i$ is the electric
charge ($\pm 1$) of particle $i$. 
The sum is over all tracks in the jet excluding the $\Bs$ decay products, 
since the latter contain no information on whether the candidate meson
was produced as a $\Bs$ 
or $\Bsbar$ and would only dilute the information from the fragmentation 
tracks. The optimal value of $\kappa$ was found to be 0.4 as 
in~\cite{cp}.
%, this value optimizes the weights given to the fragmentation tracks' charges.

The fragmentation kaon tag is an attempt to identify the kaon containing the 
$\overline{\mathrm{s}}$ quark that was produced in the fragmentation process in
association with the s quark which is part of the $\Bs$. This kaon was 
selected as follows:
\begin{itemize}
\item The probability that the measured $\dEdx$ of the track is
  consistent with the kaon hypothesis is greater than 1\%.
\item The measured $\dEdx$ of the track is lower than the expected
  value for pions of that momentum by at least one standard deviation
  of the  measurement. 
\item The measured $\dEdx$ of the track is higher than the expected
  value for protons of that momentum by at least one standard
  deviation. 
\item The track is not identified as a lepton (as in Section~\ref{sec:selection}).
\item The distance of closest approach in $r\phi$ of the track to the
  $\ee$ interaction vertex is smaller than 2~mm.
\end{itemize}
When two tracks with the same reconstructed charge satisfied these
 requirements, the event was tagged using that charge. When the
 two tracks' charges were different, or when three or more tracks
 satisfied those  requirements, the event was not given a
 fragmentation kaon tag. The latter scenario is limited to less than
 5\% of the tagged events.

\begin{figure}[htb]
\centering
\epsfxsize=13cm
\begin{center}
    \leavevmode
    \epsffile{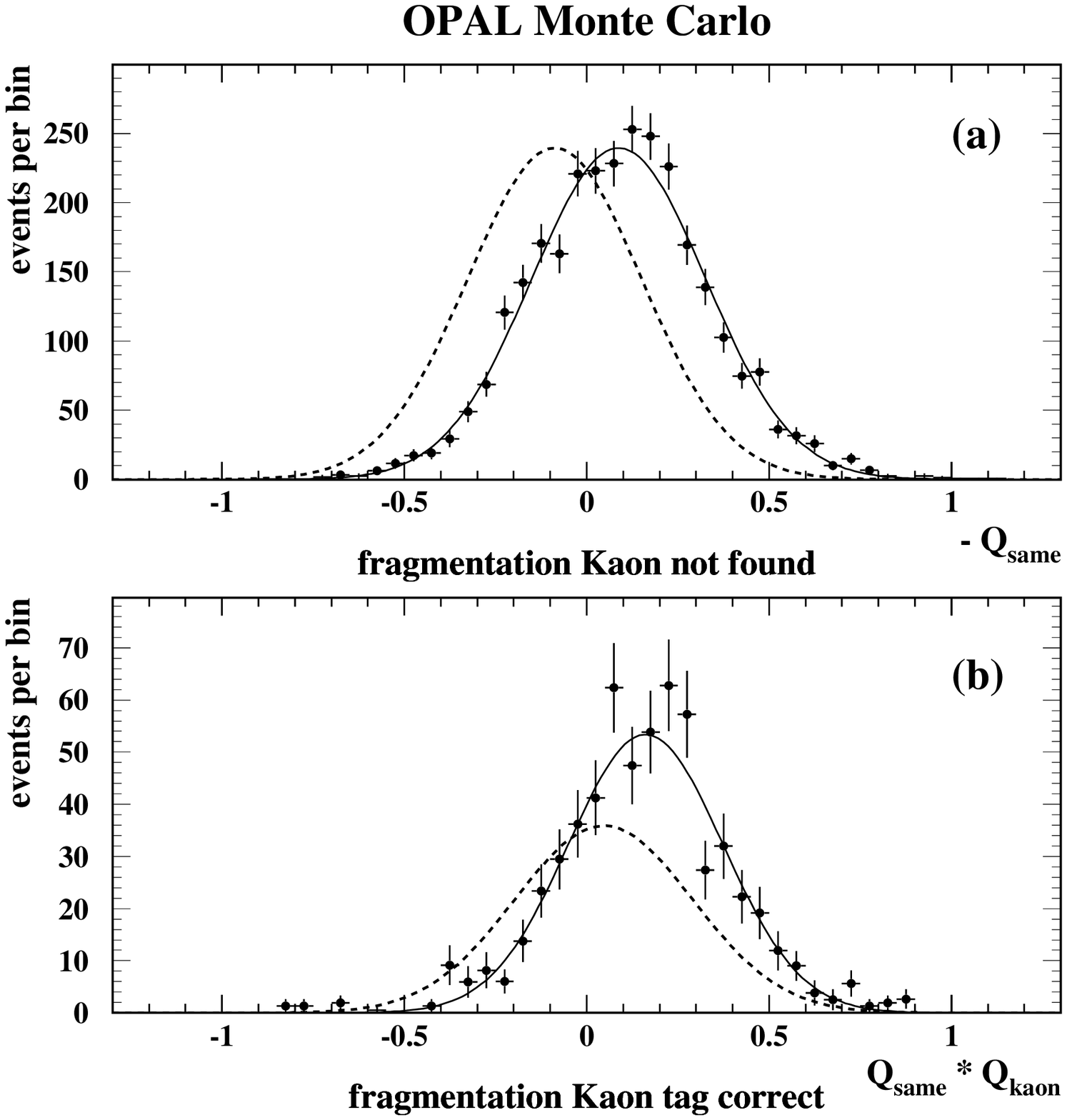}
\end{center}
\vspace{-5 mm}
\addFig{Jet charge distributions in candidate hemisphere} 
\Caption{ 
Distribution of $\Qsame$ for $\Bs$ candidates in simulated events.
Points with error bars show the number of events, the solid lines show
the Gaussian fits used in the analysis and the broken lines show the
relevant competing hypothesis.
The top plot shows events without a fragmentation kaon tag, where 
the competing hypothesis is the  
%that $Q_{same}$ arises from the 
opposite candidate flavor.
The lower plot shows events where the fragmentation kaon tag
indicates the correct candidate flavor, where the competing
hypothesis is the opposite candidate flavor and fragmentation kaon
mistag. 
Note that the competing hypothesis has a lower overall
probability since the fragmentation kaon tag's purity is more than 50\%.
%The lower plot  is of events where the fragmentation kaon tag
%indicates the wrong candidate flavor.
}
\label{fig:ftgaussame}
\end{figure}

When no fragmentation kaon was tagged, the jet charge was converted to
a probability using the Bayesian formula:
\begin{equation}
  \pr (\Bs|\Qsame)=\frac{ \pr (\Qsame|\Bs)}
  {\pr (\Qsame|\Bs)+\pr (\Qsame|\Bsbar)}~,
\end{equation}
where $\pr (\Bs)$ is
the probability of the candidate being a $\Bs$ and not a $\Bsbar$, 
and $\pr (\Qsame|flavor)$ is a Gaussian probability density
describing the $\Qsame$ distribution conditioned by the candidate's true
production flavor, as obtained from a fit to signal Monte Carlo.
This formula uses the fact that the $a \, priori$ probabilities
 of both flavors are one half.
The separation between the two competing hypotheses is shown in
Figure~\ref{fig:ftgaussame}a.

When an additional fragmentation tag was found, the jet charge and the
additional tag were converted to a probability
using the following Bayesian formulae:
\pagebreak[3]
%$$ \pr \left( \Bs|Q_{\mbox{same}},T=^{\Bs}_{\Bsbar} \right)=
\begin{eqnarray}
& \pr \left(
  \Bs|\Qsame,T={\Bs}\right) & = 
\frac{\pr(\Qsame,T={\Bs}|\Bs)}
{\pr (\Qsame,T={\Bs}|\Bs)+ \pr (\Qsame,T={\Bs}|\Bsbar)} \nonumber\\
& & =\frac{\pr \left(\Qsame|{\mathrm tag} \right) 
\cdot \pr({\mathrm tag})}
{\pr(\Qsame|{\mathrm tag},\Bs)\cdot \pr({\mathrm tag}) +
 \pr(\Qsame|{\mathrm mistag},\Bsbar)\cdot \pr({\mathrm mistag})}
 \nonumber \\
& \pr \left(
  \Bs|\Qsame,T={\Bsbar} \right) & = 
\frac{\pr(\Qsame,T={\Bsbar}|\Bs)}
{\pr (\Qsame,T={\Bsbar}|\Bs)+ \pr (\Qsame,T={\Bsbar}|\Bsbar)}\\
& & =\frac{\pr \left(\Qsame|{\mathrm mistag} \right) 
\cdot \pr({\mathrm mistag})}
{\pr(\Qsame|{\mathrm mistag},\Bs)\cdot \pr({\mathrm mistag}) +
 \pr(\Qsame|{\mathrm tag},\Bsbar)\cdot \pr({\mathrm tag})}~, \nonumber 
%& \pr \left(
%  \Bs|Q_{\mbox{same}},T=^{\mathrm{B_s}}_{\overline{\mathrm{B_s}}}
%\right) & = \frac{\pr(Q_{\mbox{same}},T|\Bs)}
%{\pr (Q_{\mbox{same}},T|\Bs)+ \pr (Q_{\mbox{same}},T|\Bsbar)} \\
%& & =\frac{\pr \left(Q_{\mbox{same}}|^{tag}_{mistag} \right) 
%\cdot \pr(^{tag}_{mistag})}
%{\pr(Q_{\mbox{same}}|tag)\cdot \pr(tag) +
% \pr(Q_{\mbox{same}}|mistag)\cdot \pr(mistag)}~, \nonumber 
\end{eqnarray}
where T is the flavor indicated by the kaon tag, 
and $\pr(\Qsame|^{\mathrm tag}_{\mathrm mistag},\Bs)$ are two Gaussian
probability distributions fitted on
% signal Monte Carlo 
simulated $\Bs$ decays
for the case when T indicates the correct flavor
 and for the case when the wrong flavor is indicated, and 
$\pr(\Qsame|^{\mathrm tag}_{\mathrm mistag},\Bsbar) = 
 \pr(-\Qsame|^{\mathrm tag}_{\mathrm mistag},\Bs)$.
 Again use was made of the fact that the $a \, priori$ probabilities
 of both flavors are one half. 
A comparison of the two competing hypotheses is shown in
Figure~\ref{fig:ftgaussame}b. 

The $\Qsame$ distribution is not necessarily charge symmetric
because of detector effects causing differences in the reconstruction
of positively and negatively charged tracks.
These effects are caused by the material in the detector
and the Lorentz angle in the jet chamber.
They were removed by subtracting an offset from the $\Qsame$
value before using it to tag the candidate's production flavor and
before parameterizing $\pr(\Qsame|^{\mathrm tag}_{\mathrm mistag},\Bs)$.
The small $\Qsame$ offset was determined from simulated signal events,
since no pure sample of fully reconstructed signal decays is available
from the data. This procedure gains support from the agreement between
the $\Qopp$ offset values calculated from simulation and data in
Section~\ref{sec:MTopp}.
After subtracting the offset, the simulated $\Qsame$
distribution is charge symmetric. 
The $\Qsame$ offset was found to be $0.006 \pm 0.004$, where the error
is from limited Monte Carlo statistics.

%       %       %        %       %       %       %       %       %       %
\subsection{The jet opposite the $\bsym \Bs$ candidate}
%       %       %        %       %       %       %       %       %       %
\label{sec:MTopp}

Flavor anticorrelation between the two hemispheres allows the use
of the b flavor in the hemisphere opposite the candidate to tag
the candidate's production flavor. The b flavor in that hemisphere was
tagged using the jet charge of the highest energy jet it contains,
and, where available, the charge of a track identified as a lepton
from semileptonic b decay.

The jet charge in the highest energy jet opposite the $\Bs$ candidate, 
$\Qopp$, was calculated in the similar way to
$\Qsame$, except 
that here the sum included all the particles in the jet and 
the optimal value of $\kappa$ was found to be 0.5 as in~\cite{cp}.
This value of  $\kappa$ optimizes the weight given to the
fragmentation tracks' charges relative to the weight given to the
decay tracks' charges.

A lepton in the opposite hemisphere was selected as follows:
\begin{itemize}
\item The track is identified as a lepton as in Section~\ref{sec:selection}.
% reference a citation about NN7 (cited text similar to NN5 text, NN5 used before)
\item Momentum greater than $2~\GeVc$.
\item Transverse momentum greater than $0.8~\GeVc$ with respect 
to the jet axis.
\item It must not be identified as arising from photon
  conversion.
\end{itemize}
When two tracks with the same reconstructed charge satisfied these
 requirements, the event was tagged using that charge. When the two tracks'
charges were different, or when three or more tracks satisfied those
 requirements, the event was not given an opposite lepton tag.

\begin{figure}[tb]
\centering
\epsfxsize=13cm
\begin{center}
    \leavevmode
    \epsffile{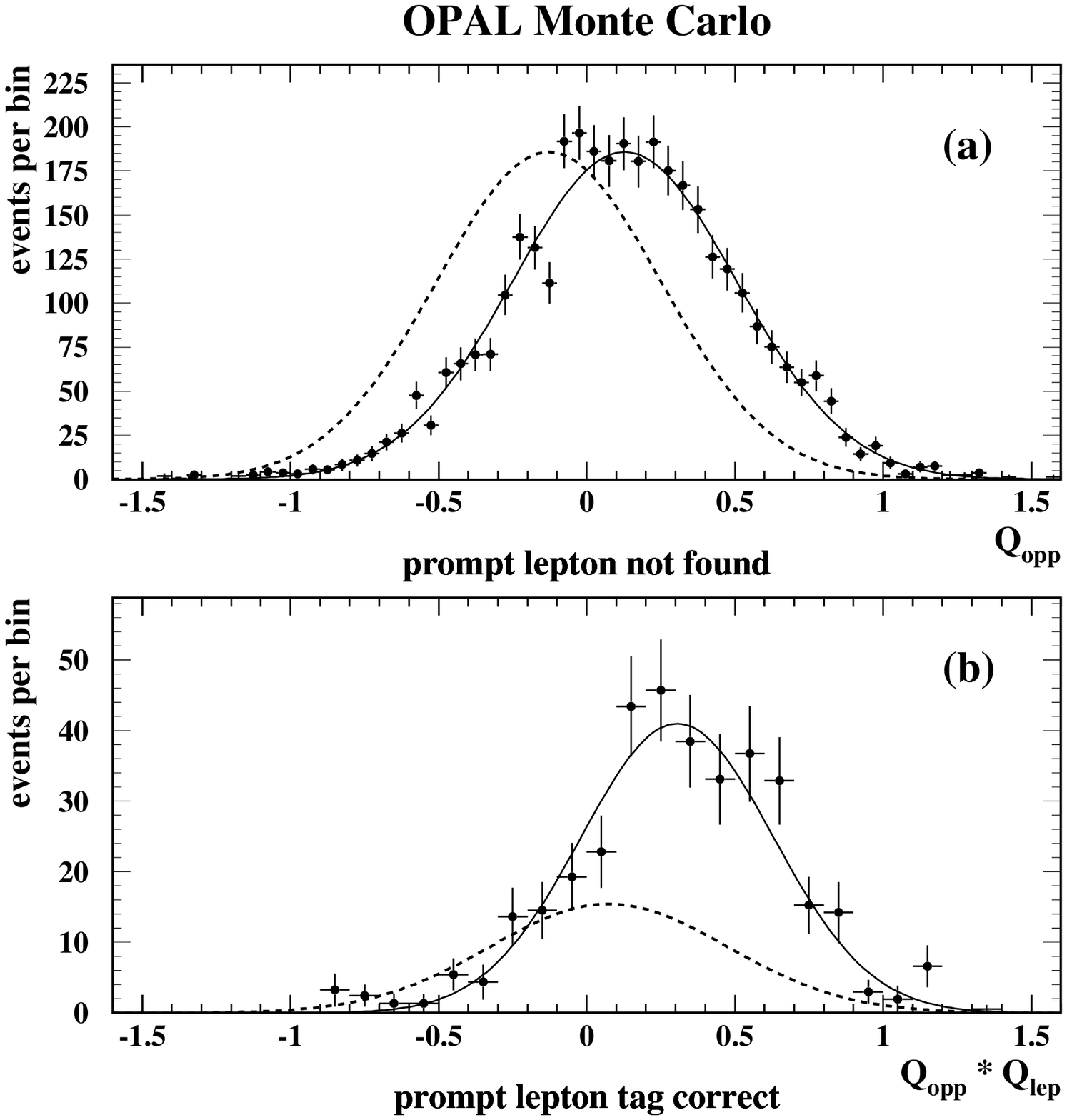}
\end{center}
\vspace{-5 mm}
\addFig{Jet charge distributions in opposite hemisphere} 
\Caption{ 
Distribution of $\Qopp$ for $\Bs$ candidates in simulated events.
Points with error bars show the number of events, the solid lines show
the Gaussian fits used in the analysis and the broken lines show the
relevant competing hypothesis.
The top plot shows events without an opposite lepton tag, where 
the competing hypothesis is the other candidate flavor.
The lower plot shows events where the opposite lepton tag
indicates the correct candidate flavor, where the competing
hypothesis is the other candidate flavor and opposite lepton
mistag. 
Note that the competing hypothesis has a lower overall
probability since the opposite lepton tag's purity is more than 50\%.
%The lower plot  is of events where the opposite lepton tag
%indicates the wrong candidate flavor.
}
\label{fig:ftgausopp}
\end{figure}

The jet charge and the lepton tag of the opposite hemisphere 
were converted to a probability using the same method as in
Section~\ref{sec:MTcand} for the hemisphere containing the $\Bs$
candidate. A comparison of the two competing hypotheses is shown in
Figure~\ref{fig:ftgausopp}. 

As described in Section~\ref{sec:MTcand} for $\Qsame$, the 
$\Qopp$ distribution is not charge symmetric.
The $\Qopp$ offset was determined using
a large sample of b tagged inclusive lepton events selected from data.
The resulting value of the $\Qopp$ offset agrees well with the values
derived from simulated signal events, from a large simulated sample of
b tagged inclusive lepton events, and from~\cite{cp}.
After subtracting the offset, the simulated $\Qopp$
distribution is charge symmetric. 
The $\Qopp$ offset was found to be $0.0138 \pm 0.0020$, where the error
is from the limited statistics of the selected data sample.

%       %       %        %       %       %       %       %       %       %
\subsection{Mixing tag results}
%       %       %        %       %       %       %       %       %       %
\label{sec:mtres}
The procedure described above attempts to assess the probability
that an event underwent mixing. Being based on Gaussian approximations
of the jet charge distributions, this raw mixing tag, $x$, is
therefore only an approximation of the true mixing probability.
A calculation of the
events likelihood demands that we calibrate the mixing tag, $M$, by
quantifying its deviation from a true probability.

\begin{figure}[tb]
\centering
\epsfxsize=11cm
\begin{center}
    \leavevmode
    \epsffile{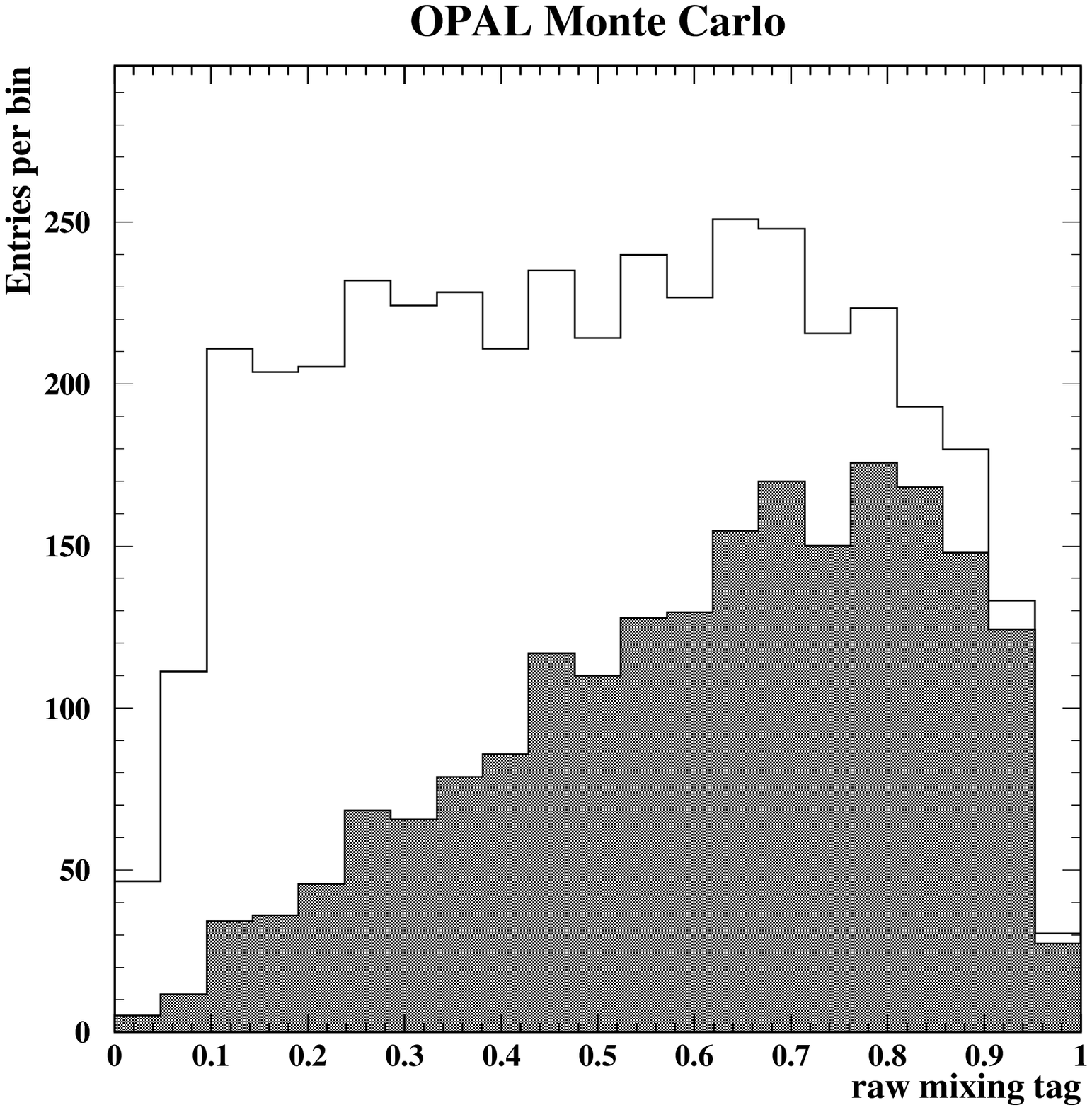}
\end{center}
\vspace{-5 mm}
\addFig{Distribution of raw mixing tag in signal events}
\Caption{ 
Distribution of the raw mixing tag in simulated signal events.
The light histogram shows the distribution of the raw mixing tag, 
the dark histogram shows the contribution of mixed events.
}
\label{fig:mtdist}
\end{figure}

The average mixing tag for simulated signal events, in which half the
decays were mixed, was found to be consistent with one half. Furthermore
the difference between this average and the true average (0.5)
is much smaller than the typical systematic uncertainties on the
mixing tag, and is neglected in this analysis. The distribution of the
mixing tag in simulated signal events is shown in Figure~\ref{fig:mtdist}.

\begin{figure}[tb]
\centering
\epsfxsize=11cm
\begin{center}
    \leavevmode
    \epsffile{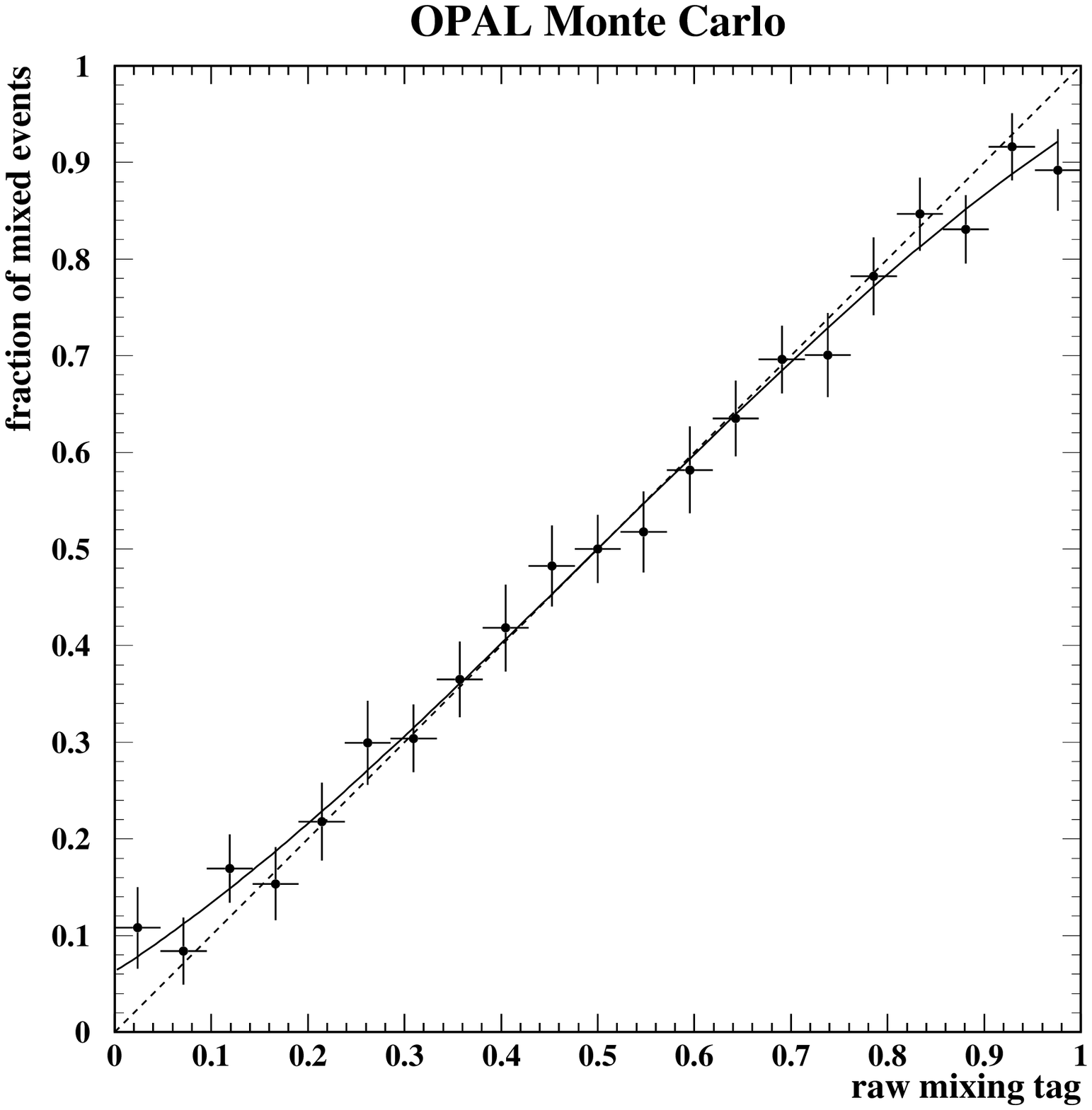}
\end{center}
\vspace{-5 mm}
\addFig{Systematic errors on mixing tag}
\Caption{ 
Deviation of mixing tag from a true probability.
The points with the error bars show simulated events, the solid line
is a fit to these points as
described in the text and the dotted line is $M=x$, representing a true
probability requiring no correction. The error bars include systematic
errors described in Section~\ref{sec:system}.
Both charge conjugated cases were added to maximize the 
statistical significance.
}
\label{fig:mtsys}
\end{figure}

The deviation of the mixing tag from the $a\, posteriori$ probability is
parameterized with the parameters $\ftmid$ and $\ftend$ (used in the
oscillation fit Section \ref{sec:like}), which quantify the deviations
at $x=0.75$ and $x=1.0$ respectively.
% as:
%\begin{equation}
% M=\min \left(1,\max \left(0,\half+a\left(x-\half \right)
% +2(1-a)\left(x-\half \right) \cdot \left\vert x-\half \right\vert+
% b\left(x-\half \right)^3 \right) \right)~,
%\end{equation}
%where $x$ is the probability estimated from the mixing tag,
%$M$ is the
%probability used in the fit after correcting according to the
%deviation.
%This parametrization was chosen for the following reasons:
%asymmetry around the $x=\half$ point and the ability to control the
%behavior around the $x=0,1$ and $x=\half$ points with small correlation.
%The parameters $\ftmid$ and $\ftend$ used in the oscillation fit
%(Section \ref{sec:like}), quantify the deviation from the $M=x$ line
%and are given by  $a=1+8\ftmid$ and $b=8\ftend$.
%
The 
% fitted deviation of the combined tag from the $a\, posteriori$ probability 
resulting fit
is shown in Figure~\ref{fig:mtsys}, and the fitted 
uncertainty is treated as a systematic error.

%       %       %        %       %       %       %       %       %       %
\subsubsection {Mixing tag behavior in the combinatorial background}
%       %       %        %       %       %       %       %       %       %
\label{sec:combMT}
%The mix tag presented in section ~\ref{sec:MT} was optimized to
%tag signal decays.

To calculate the likelihood of an event originating from
combinatorial background, we need to know the behavior
of the mixing tag in combinatorial background events.
Two effects which influence this behavior were found using 
Monte Carlo: a different mixing tag distribution, and
oscillations of the background. These effects were identified in
particular subsamples of the
combinatorial background. The definition, abundance and behavior of
those subsamples are as follows.

Combinatorial background events which originated in $\bbbar$ events
can exhibit oscillatory behavior if the decaying meson is a $\Bd$ or
a $\Bs$. Two rates of oscillation were found in simulated
combinatorial background events:

Oscillation with the same rate as the simulated signal
oscillation rate arises primarily when the decay is truly through a
$\Bs$ and only one of the $\Ds$ decay products is
misidentified.
The proper decay time reconstruction for these simulated events was
less accurate than for signal events, the width (RMS) of the difference
between the true time and the expectation value of the reconstructed
time distribution is bigger by $25\pm7\%$ for those events.
%$\fixme$ need to rerun probber to update these numbers,
% were created with older definition which is no longer needed and
% only estimated from dl (gave 21%) behavior so far.
The mixing tag for these simulated events was essentially
the same as for simulated signal events.

It was found that oscillation with the same rate as the $\Bd$
oscillation rate arises primarily when 
%the decay is through a $\Bd$,
the decay products of a  $\D^+$ or $\D^0$ meson produced from a $\Bd$
were reconstructed as a $\Ds$ meson, and at most one of the D meson
decay products was misidentified.
A typical decay chain for this channel is $\Bd \to \D^{(*)}\lp\nu\X$,
 which is similar to the signal decay $\Bs \to \D_s\lp\nu\X$.
The mixing tag performance and decay time reconstruction for these
simulated events were consistent with that for simulated signal
events.
%This simulation included further oscillations of $\D^0$ mesons,
%produced in $\B$ decays, that were reconstructed as $\Ds$ mesons. 
%see ~/bs/text/momentumNumbers

Monte Carlo studies have shown that the fraction, $\fOCB$, of
combinatorial background which oscillates depends on whether
the channel contains a $\phi$, both for $\Bs$ and $\Bd$ fractions.
Therefore we use four parameters in the fit, $\fOCBdphi$, $\fOCBdno$,
$\fOCBsphi$ and $\fOCBsno$, that give the $\Bs$ and $\Bd$ fractions
for $\phi$ and No-$\phi$ channels.

The distribution of the mixing tag for the non-oscillating
combinatorial background is different from that for signal, hence the
mixing tag not only indicates mixing but also contributes information
as to whether the event is a signal event. 
Our use of this information is described in Section~\ref{sec:physMT}.
% We derive a probability, $\fmixi$, given only
% the mixing tag, that an event
% is from the combinatorial background. The shape of $\fmixi$ as a
% function of the mixing tag was fitted on simulated events, and the
%% resulting deviation from $\fmixi = 0.5$ was described by a single
%% parameter, $g$.
% fitted uncertainty of the shape was used as a systematic uncertainty.

% The bias of the mixing tag is needed to derive the expected average
% mixing tag (Section~\ref{sec:chemix}).
Biases were found mainly on non-$\bbbar$ events.
%by enlarging cuts: cc and bb biases significant (bb small)
While uds
%${\mathrm{\overline{u}}}$,
%d${\mathrm{\overline{d}}}$ and s${\mathrm{\overline{s}}}$ 
events tend
to be tagged as mixed, heavier flavor events have a statistically
significant tendency to be tagged as unmixed.
These tendencies partially cancel out, leaving an overall
bias of the mixing tag on non-oscillating combinatorial background
events that was found to be $\ccbias = -0.030 \pm 0.009$.

%       %       %        %       %       %       %       %       %       %
\subsubsection {Mixing tag behavior in the $\bsym {\Ds \lp}$ background}
%       %       %        %       %       %       %       %       %       %
\label{sec:physMT}

The mixing tag for simulated $\Dsl$ background events of type (a)
(as defined in Section~\ref{sec:phys-bg}) was
essentially opposite that for simulated signal events, as expected 
from the decay chain. In addition,
the distribution of the mixing tag for $\Ds \lp$ background from $\Bu$
decay was found to be different from that for signal, while for 
$\Ds \lp$ background from $\Bd$ decay the distribution of the mixing
tag was consistent with that for signal.
Hence the mixing tag not only indicates mixing, but also contributes
information as to whether the event
is from the signal, from the combinatorial background
(Section~\ref{sec:combMT}), or from the $\Bu \to \Ds \lp$ background.
We derived a probability, $\fmixi$, given only
the mixing tag, that an event is from the combinatorial background,
and similarly a probability, $\fphysi$, that it is from the  
$\Bu \to \Ds \lp$ background.
Using simulated events we fitted $\fmixi$ and $\fphysi$ as
functions of the mixing tag, and the
% resulting deviation from $\fmixi = 0.5$ was described by a single
% parameter, $g$.
% fit as 2nd and 3rd degree polinomials, respectively.
fitted uncertainties were used as a systematic uncertainties.

%%%%%%%%%%%%%%%%%%%%%%%%%%%%%%%%%%%%%%%%%%%%%%%%%%%%%%%%%%%%%%%%%%%%%%%%
\section{Oscillation fit}
%%%%%%%%%%%%%%%%%%%%%%%%%%%%%%%%%%%%%%%%%%%%%%%%%%%%%%%%%%%%%%%%%%%%%%%%
%\label{sec:oscfit}
%
%       %       %        %       %       %       %       %       %       %
%\subsection{Likelihood function}
%       %       %        %       %       %       %       %       %       %
\label{sec:like}

The likelihood, ${\cal L}$, for observing a particular decay length,
$\li$, of candidate $i$, and a particular mixing tag, $\Mi$, may be
parameterized in terms of the candidate's
decay length error, $\sigLi$, its calculated $\Bs$
momentum spectrum (Section ~\ref{sec:momest}), $\Bi$,
and a probability, $\fmassi$, that it arises from
combinatorial background.
$\fmassi$ is determined as a function of the observed invariant mass of
this candidate from the fit to the invariant mass spectrum shown in
Figure~\ref{fig:Dsmass}.

An event's likelihood is found by summing over all the possible event
types (i.e. signal, the two $\Ds \lp$ background modes, the two oscillating
combinatorial background modes, and regular combintorial background).
For each event type we assign a probability that it is an
event of this type, ${\pr}({\mathrm type})$, and the likelihood if the event
is of that type, ${\cal L}_i^{\mathrm type}$:
\begin{equation}
  \Lall(\li, \Mi \mid \sigLi, \Bi, \fmassi) =
   \sum_{\mathrm event\ type}
    {\pr}\Big({\mathrm type} \Big| \fbgi (\fmassi,\fmixi), \fphysi \Big) \cdot
    {{\cal L}_i^{\mathrm type}}(\li, \Mi \mid \sigLi, \Bi)~,
\end{equation}
where $\fbgi$ is the estimated probability that the event arises from
combinatorial background based on the invariant mass (through
$\fmassi$) and on the mixing tag (through $\fmixi$), and $\fphysi$ is
the estimated probability, based on the mixing tag, that the event is
from the $\Bu \to \Ds \lp$ background.

The form of the likelihood function for signal events is given by
 the convolution of three terms: a term describing the probability of
 the mixing tag and the true decay length given the true momentum, the
 calculated momentum 
 distribution, and a Gaussian resolution function with width equal
 to the decay length error (corrected as described in
 Section~\ref{sec:dlest}). 
 This can be expressed as:
\begin{eqnarray}
\Lsig(\li, \Mi \mid \sigLi, \Bi)  & = &  \int_0^\infty \ud \lit \\
& & \int_0^{\infty} \ud\pb \ \   
    {\cal G}(\li \mid \lit,\sigLi) \ \,
    \Bi(\pb) \ \,
    {\cal P}_{\mathrm signal}(\lit, \Mi \mid \pb)~, \nonumber 
\end{eqnarray}
where the function ${\cal G}$ is a Gaussian
function that describes the probability to observe a decay length,
$\li$, given a true decay length $\lit$ and the estimated measurement
uncertainty $\sigLi$.  
$\Bi(\pb)$ is the probability of a particular $\Bs$
momentum. 
% The true decay distribution term combines the contributions of mixed and
%  unmixed decays, weighting them according to the mixing tag.
${\cal P}$ is the
probability for a given $\Bs$ to decay at a distance
$\lit$ from the $\ee$ interaction vertex with a mixing tag
$\Mi$. This function is given by:
% technically this is inaccurate, as we canceled out the apriori
% mixing tag distribution term (which is really part of P(type|..)
\begin{eqnarray}
\label{equ:psig}
  {\cal P}_{signal} (\lit, \Mi \mid \pb) & = & \frac{\mb}{\pb} \cdot
    \frac{ \exp (-t/\taubs)}{\taubs} \cdot \\
& & \left( \frac{1+{\cal A}\cdot \cos(t \dms)}{2} \cdot (1-\Mi) + 
           \frac{1-{\cal A}\cdot \cos(t \dms)}{2} \cdot \Mi \right)~, \nonumber 
\end{eqnarray}
where $\taubs$ is the $\Bs$ lifetime, ${\cal A}$
is the fitted amplitude of the oscillation \cite{moser}
and the $\Bs$ candidate's true proper decay time, $t$,
is given by Equation~\ref{equ:propertime}.
% $t=\frac{\lit \cdot m_{\mathrm {\tiny B_s}}} {\pb}$
% % p_{\mathrm  {\tiny B}}
% is the $\Bs$ candidate's true proper decay time, given its
% true momentum, mass, and decay length.

For the $\Ds \lp$ background events from $\Bd$ decay, the likelihood is
similar except that the $\Bd$ lifetime was used to obtain:
\begin{eqnarray}
\label{equ:posc}
  {\cal P}_{\mathrm osc} (\lit, \Mi \mid \pb) & = &
    \frac{\mb}{\pb} \cdot \frac{ \exp (-t/\tbd)}{\tbd} \cdot \nonumber \\
& & \left( \frac{1 + \cos(t \dmd)}{2} \cdot (1-\Mi) + 
           \frac{1 - \cos(t \dmd)}{2} \cdot \Mi \right)~,
\end{eqnarray}
where for decay mode (a) we need to replace $\Mi$ with $1-\Mi$.
For the $\Ds \lp$ background events from $\Bu$ decay, the likelihood
is simpler, and contains an exponential decay term with the $\Bu$
lifetime weighted by the candidate's probability to be unmixed.

                         %%% COMBINATORIAL BG %%%
 
The combinatorial background was divided into several types according
to oscillatory behavior, with most of the combinatorial background
events being of the non-oscillating type.
The function used to parameterize the reconstructed decay length
distribution of
this background is the sum of a positive and a negative
exponential, convoluted with the same boost function as the signal
and a Gaussian resolution function. This can be expressed as:
\begin{equation}
 {\cal P}_{comb} (\li \mid \tbgp,\tbgn,\fp,\pb) = 
\left\{ \begin{array}{ll}
 \fp\,\frac{\mb}{\tbgp\pb}\,
        \exp\left[\frac{-\,\li \cdot \mb}{\tbgp\pb}\right]
  & \textrm{if $\li \ge 0$}\\
 (1-\fp)\,\frac{\mb}{\tbgn\pb}\,
      \exp\left[\frac{-\,(-\li) \cdot \mb}{\tbgn\pb}\right]
  & \textrm{if $\li < 0$}~.
 \end{array} \right.
\end{equation}
The fraction of background with positive lifetime, $\fp$, as well as
the characteristic positive and negative lifetimes of the background,
$\tbgp$ and $\tbgn$, were obtained from a fit to the sideband region.
The resulting value and their uncertainties were used to constrain the
background lifetime parameters in the oscillation fit. 
The background parameters
were fitted separately for the hadronic and semileptonic $\Ds$ decay
channels, as in~\cite{lifetimepaper}.
%outdated draft1 fit:
%For the hadronic $\Ds$ decay channels the sideband fit yielded 
%$\fp = 1.02 \pm 0.06$ indicating that the negative exponential
%lifetime distribution is redundant for these channels. Fitting only a
%positive exponential decay yielded $\tbgphad = 0.75 \pm 0.09 \ps$ and
%a better fit.
%The fit to the semileptonic $\Ds$ decay events in the sideband region
%yielded $\fplep = 0.94 \pm 0.06$, $\tbgplep = 1.51 \pm 0.28 \ps$ and
%$\tbgnlep = -0.7 \pm 0.5 \ps$.
The lifetime behavior of the hadronic $\Ds$ decay channels' sidebands
was best described by fitting only a positive exponential decay.
% with
% the decay length resolution increased, relative to the signal, by
% $\dlecomb = 1.78 \pm 0.26$. 
The lifetime behavior of the semileptonic $\Ds$ decay channel's sideband
was best described by fitting both exponential terms.
% without
% increasing the decay length resolution relative to signal.

For the semileptonic channels, the background which include a real
$\phi$ not from a $\Ds$ is treated as combinatorial background.
For the oscillating types of combinatorial background we used the
following: for background oscillating at the $\Bd$ frequency we
used Equation~\ref{equ:posc}, while for background oscillating at the
$\Bs$ frequency we used Equation~\ref{equ:psig}.

%%%%%%%%%%%%%%%%%%%%%%%%%%%%%%%%%%%%%%%%%%%%%%%%%%%%%%%%%%%%%%%%%%%%%%%%
\section{Results of oscillation fit}
\begin{figure}[htbp]
\centering
\epsfxsize=17cm
\begin{center}
  \leavevmode
  \epsffile{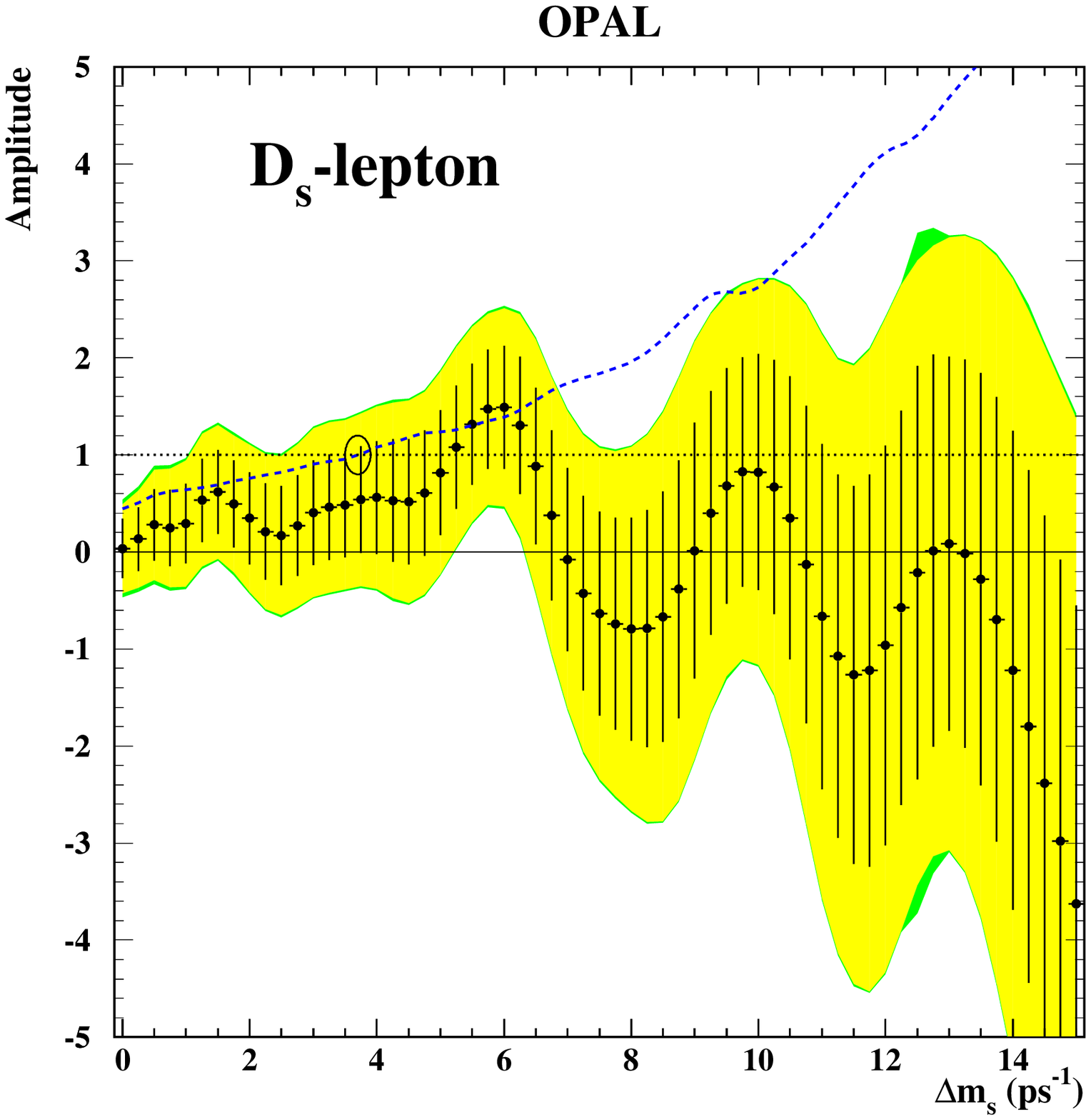}
\end{center}
\vspace{-5 mm}
\addFig{Results of amplitude fit}
\Caption{ 
Measured $\Bs$ oscillation amplitude as a function of $\dms$ for this
analysis. 
The error bars represent the 1$\sigma$ statistical uncertainties.
The shaded bands show the $\pm 1.645 \sigma$ region, with and without
including systematic effects. The values of $\dms$ where the shaded
region lies below the ${\cal A}=1$ line are excluded \clnf.
The dashed line is 1.645$\sigma$ used to determine the experimental
sensitivity, which is indicated by the circle. 
}
\label{fig:amp}
\end{figure}

%%%%%%%%%%%%%%%%%%%%%%%%%%%%%%%%%%%%%%%%%%%%%%%%%%%%%%%%%%%%%%%%%%%%%%%%
\label{sec:resfits}
The results of the amplitude fit to the selected events are shown in
Figure~\ref{fig:amp}, including the systematic uncertainties
(Section~\ref{sec:system}).
An amplitude peak is evident at $\dms=6.0\psm$,
above the experimental sensitivity, but nevertheless it seems
inconsistent with an amplitude of zero with a significance of 2.35 sigma
(including systematic uncertainties).
% 2.348, including systematics 
The current combined world lower limit is $\dms > 14.4\psm$; this
leads us to interpret this peak as a statistical fluctuation.
At low frequencies the fitted amplitude quickly rises above the 
$A = 0$ line, and therefore after taking into account all systematic
uncertainties described in Section~\ref{sec:system}, this analysis can
only set a weak lower limit of $\dms > \limit\psm$ \clnf.

%Part of the positive fitted amplitudes for low values of $\dms$
%%can be attributed to the total mixing tag as described in
%section~\ref{sec:chemix}, since at low $\dms$ values ($\dms$ less than
%about $2.0~\psm$) $\chi_s$ is significantly less than half.
%Though the deviation of the total mixing tag
%from the total mixing tag expected for rapid oscillations is
%statistically insignificant, it does not imply that its influence on
%the resulting amplitude is negligible. At $\dms=0~\psm,$ the amplitude
%determines the predicted average mixing tag of signal events. An
%amplitude of $0.09$ is necessary to account for the observed deviation of
%the total mixing count. This is close to the amplitude actually fitted at
%$\dms=0~\psm$, which is $0.12\pm0.28$, confirming the interpretation of this
%part of the positive amplitudes as a result of the total mixing count
%fluctuation.

%

%%%%%%%%%%%%%%%%%%%%%%%%%%%%%%%%%%%%%%%%%%%%%%%%%%%%%%%%%%%%%%%%%%%%%%%%
\section{Systematic uncertainties}
%%%%%%%%%%%%%%%%%%%%%%%%%%%%%%%%%%%%%%%%%%%%%%%%%%%%%%%%%%%%%%%%%%%%%%%%
\label{sec:system}
 
The systematic uncertainties on the $\Bs$ oscillation amplitude, 
$\sysErr$, are calculated, using the prescription of~\cite{moser}, as:
\begin{equation}
%  \sysErr & = & \frac {\sysErr^+ - \sysErr^-} {2} \nonumber \\
  \sysErr = {\cal A}^{\mathrm {new}}  - {\cal A}^{\mathrm {nominal}} +
            ( 1 - {\cal A}^{\mathrm {nominal}} ) 
            \frac
              {\sigma^{\mathrm {new}} - \sigma^{\mathrm {nominal}}}
              { \sigma^{\mathrm {nominal}}},
\end{equation}
where the superscript ``nominal'' refers to the amplitude value,
${\cal A}$, and statistical uncertainty, $\sigma$, obtained using the
nominal values of the various parameters, and the ``new'' refers to
the new values obtained when a single parameter is increased or
decreased by its uncertainty and the fit is repeated. 
The systematics shown are an average of the effects of the increment and the
decrement.
The nominal values and errors used are given in
Table~\ref{tab:pars}, their description follows:

\begin{description}
\item[$\Ds\lp$ background fraction:]
 The fractions of the two modes of $\Ds \lp$ background out of events in the
signal peak (in the case of the semileptonic $\Ds$ decay channel it is done
after subtraction of $\fother$), $\fphysa$, $\fphysb$, and the
fraction of $\Bd$ decays in each mode, $\fraca$ and $\fracb$ were
calculated in Section~\ref{sec:phys-bg}. 
Due to the dependance of $\fphysa$ and  $\fphysb$ on the $\Bs$
production fraction $f_s$, the uncertainty on the value of $f_s$ is a
source of systematic uncertainty.

\item[B meson lifetimes, oscillation and production fractions:]
The world averages \cite{PDG} for the lifetime of the $\Bu$, $\Bd$
and $\Bs$ mesons, for the $\Bd$ oscillation frequency and for the
$\Bs$ production fraction were used.

\item[Oscillating combinatorial background fractions:]
The fractions of both types of oscillating combinatorial background
out of the total combinatorial background for the various $\Ds$ decay
channels:
$\fOCBdphi$,$\fOCBdno$,$\fOCBsphi$ and $\fOCBsno$ were calculated in
Section~\ref{sec:combMT}.

\item[Other background fraction:]
The fraction of other types of background specific to semileptonic $\Ds$
decays out of events in the signal peak: $\fother$ was estimated as
described in Section~\ref{sec:otherbg}.

\item[Combinatorial background lifetime:]
The combinatorial background parameters \\
$\fplep$, $\tbgplep$ and $\tbgnlep$ for semileptonic $\Ds$ decays; 
and $\tbgphad$ for hadronic $\Ds$ decays, were obtained
from a fit to sideband events (Section~\ref{sec:like}).
The statistical errors on the parameters from
the sideband fit were used as systematic uncertainties.

\item[Mixing tag behavior in signal events:]
The uncertainties on the fraction of mixed events in each bin used to
parameterize the deviation of the mixing tag from a true
probability (as described in section~\ref{sec:mtres} and shown in
Figure~\ref{fig:mtsys}) include both statistical uncertainties from the
limited Monte Carlo sample size, typically of order 0.025, and the
following systematic effects:
\begin{itemize}
\item the effect of a one standard deviation variation in the $\Qsame$ offset,
  typically of order 0.004.
\item the effect of a one standard deviation variation in the $\Qopp$ offset,
  typically of order 0.005.
\item the effect of a one standard deviation variation in the fragmentation kaon tag's
  purity, typically of order 0.007.
\item the effect of a one standard deviation variation in the opposite lepton tag's
  purity, typically of order 0.006.
\item since the limited candidate sample size in data prevents us from
  showing that the simulation and data agree on the
%  $\pr(\Qsame|^{\mathrm tag}_{\mathrm mistag},\Bs)$ and $\pr(\Qsame|{\mathrm no~tag},\Bs)$
  jet charge distributions with and without a fragmentation kaon tag,
  we take the entire effect of using those
  distributions instead of the single jet charge distribution
  as a systematic error, typically of the order of 0.03.
\end{itemize}
The fitted uncertainties on the deviation parameters $\ftmid$ and
$\ftend$ were taken as the systematic uncertainties on the mixing tag.

\item[Mixing tag behavior in background:]
The fitted uncertainties on the deviations of the distributions of the
mixing tag in combinatorial and $\Bu\to\Ds\lp$ backgrounds, relative
to its behavior in signal (Section~\ref{sec:physMT}), were taken as
additional systematic uncertainties.

\item[Decay length error correction:]
The fitted uncertainty on the decay length error correction 
(Section~\ref{sec:dlest}) was used as a systematic
uncertainty.

\item[Decay time reconstruction bias:]
The fitted uncertainties on the residual bias in the decay time
reconstruction (Section~\ref{sec:dtres}) were used as systematic
uncertainties.

\item[Detector resolution modelling:]
The resolution of the tracking detectors might affect the decay time
reconstruction and the mixing tag.
The simulated resolutions were degraded by $10\thinspace\%$ relative
to the values that optimally describe the data following the studies
in~\cite{gccel}. The analysis was repeated and the mixing tag was
found to be insensitive to this variation while the proper decay time
resolution deteriorated by $5\thinspace\%$.
This $5\thinspace\%$ uncertainty on the proper decay time
resolution was used as a bidirectional systematic uncertainty.

\end{description}

\begin{table}[hp]
 \centering
 \begin{tabular}{l|c|c|l|l|l|l}
  
  input        & nominal    &       & \multicolumn{4}{c}{contribution to $\Delta {\cal A}$ at $\dms$ =} \\
  \cline{4-7} 
  parameter    & value      & error & $0~\psm$& $5~\psm$& $10~\psm$& $15~\psm$\\
  \hline

  \multicolumn{7}{l}{\mystrut Fractions of $\Ds \lp$ background:} \\
  $\fphysa$    & 0.143      & 0.050 & +0.09   & +0.06   & +0.09   & +0.11  \\%4
  $\fraca$     & 0.619      & 0.045 & $+0.0011$ & +0.0019 & +0.004  & +0.005 \\%5
  $\fphysb$    & 0.065      & 0.035 & $-0.010$  & +0.030  & $-0.009$  & $-0.033$ \\%6
  $\fracb$     & 0.470      & 0.038 & +0.0019 & $-0.0010$ & $-0.027$  & +0.0013\\%7
  
  \multicolumn{7}{l}{\mystrut Previously measured B meson properties:} \\
  $f_s$        & 0.107      & 0.014 & $-0.030$  & $-0.028$  & $-0.03$ & $-0.030$ \\%27
  $\tbu$ & $1.653\ps$ & $0.028\ps$& +0.0005 & $-0.0004$ & $-0.013$ & $-0.0013$\\%8
  $\tbd$ & $1.548\ps$ & $0.032\ps$& +0.0021 & +0.0001 & $-0.004$  & +0.0008\\%13
  $\taubs$&$1.493\ps$ & $0.062\ps$& $-0.004$  & +0.004  & +0.016  & $-0.04$  \\%3
  $\dmd$  &$0.472\psm$&$0.017\psm$& +0.0009 & +0.0004 & $-0.0002$ & $-0.0002$\\%12
  
  \multicolumn{7}{l}{\mystrut Fractions of oscillating combinatorial background:} \\
  $\fOCBdphi$  & 0.163      & 0.030 & $-0.0033$ & $-0.004$  & $-0.015$  & $-0.018$ \\%17
  $\fOCBdno$   & 0.270      & 0.027 & $-0.005 $ & +0.0032 & +0.0002 & +0.008 \\%15
  $\fOCBsphi$  & 0.163      & 0.030 & $-0.004 $ & $-0.017$  & $-0.022$  & $-0.05$  \\%18
  $\fOCBsno$   & 0.049      & 0.013 & $-0.0029$ & +0.007  & +0.010  & +0.04  \\%16

  \multicolumn{7}{l}{\mystrut Fraction of other background:} \\
  $\fother$    & 0.135      & 0.057 & +0.006  & +0.013  & +0.010  & +0.05  \\%21
    
  \multicolumn{7}{l}{\mystrut Combinatorial background lifetime fit:} \\
  $\fplep$     & 0.92       & 0.05  & +0.0020 & +0.0024 & $-0.0016$ & +0.011 \\%9
  $\tbgplep$   &$1.40\ps$& $0.26\ps$& +0.0018 & $-0.005$ & $-0.022 $ & $-0.04$  \\%10
  $\tbgnlep$   &$0.7\ps $& $0.5\ps$ & $-0.0013$ & $-0.007$ & $-0.012 $ & $-0.024$ \\%11
  $\tbgphad$   &$0.70\ps$& $0.11\ps$& +0.014  & +0.011  & $-0.04  $ & +0.06  \\%20
  
  \multicolumn{7}{l}{\mystrut Mixing tag behavior in signal events:} \\
  $\ftmid$     & $-0.006$   & 0.012 & $-0.06$ & $-0.024$ & +0.021  & +0.005 \\%22
  $\ftend$     & $-0.051$   & 0.026 & $-0.04$ & $-0.06 $ & $-0.08$ & $-0.06$\\%23

  \multicolumn{7}{l}{\mystrut Mixing tag distributions in background:} \\
  combinatorial  &        1 & 0.18  & $-0.017$  & +0.0033 & $\phantom {-}0$ & $-0.04 $\\%24
  $\Bu\to\Ds\lp$ &        1 & 0.14  & $-0.007$  & +0.0023 & $-0.0023$ & $-0.011$\\%19
  
  \multicolumn{7}{l}{\mystrut Decay length resolution:} \\
  fitted correction & 1.478 & 0.033 & +0.0009 & +0.016  & $-0.005$ & +0.06  \\%14
  detector modelling& 1     & 0.05  & +0.0012 & +0.035  & $-0.010$  & +0.14  \\%28
  
  \multicolumn{7}{l}{\mystrut Decay time bias:} \\
  $t_{\rm bias}^0$&$-0.033\ps$ &$0.005\ps$ &+0.0008& $-0.006$ & +0.031  & +0.33  \\%25
  $t_{\rm bias}^{\rm slope}$ & 1.29& 0.16      &+0.0010& $-0.005$ & +0.009  & +0.07  \\%26
  
  \noalign{\medskip \hrule width 6.125 in}
  \multicolumn{3}{|l|}{systematic uncertainty} & 0.12 & 0.11 & 0.15  &
  0.41 \hspace {3 ex}$\vrule$ \\
  \multicolumn{3}{|l|}{statistical uncertainty}& 0.28 & 0.64 & 1.2   &
  3.0 \hspace {4.1 ex}$\vrule$ \\
  \noalign{\hrule width 6.125 in}
 \end{tabular}
 \addTable{Systematic errors}
 \Caption{
    Input parameters for the fit and their contributions to the
    systematic errors, as described in the text. Systematic errors of
    less than $5\ex{-5}$ are reported as zero.
%, reflecting the limited accuracy of their estimation. 
    For comparison the total systematic
    errors and the statistical errors are given as well, in the last rows.
 }    
 \label{tab:pars}
\end{table}

The relative importance of the various systematic uncertainties, as a
function of $\dms$, is shown in Table~\ref{tab:pars}. For all $\dms$
values, the total systematic uncertainty is small compared to the
statistical uncertainty. At low $\dms$ the most important systematic 
contributions are from the uncertainties on the behavior of the
mixing tag in the signal, while at high $\dms$ the most important
systematic contributions are from the uncertainties on the decay time
reconstruction bias.

%%%%%%%%%%%%%%%%%%%%%%%%%%%%%%%%%%%%%%%%%%%%%%%%%%%%%%%%%%%%%%%%%%%%%%%%
\section{Checks of the method}
%%%%%%%%%%%%%%%%%%%%%%%%%%%%%%%%%%%%%%%%%%%%%%%%%%%%%%%%%%%%%%%%%%%%%%%%
\label{sec:checks}

%       %       %        %       %       %       %       %       %       %
\subsection{Fitted decay length distribution and lifetime fit}
\begin{figure}[htbp]
\centering
\epsfxsize=15cm
\begin{center}
    \leavevmode
    \epsffile{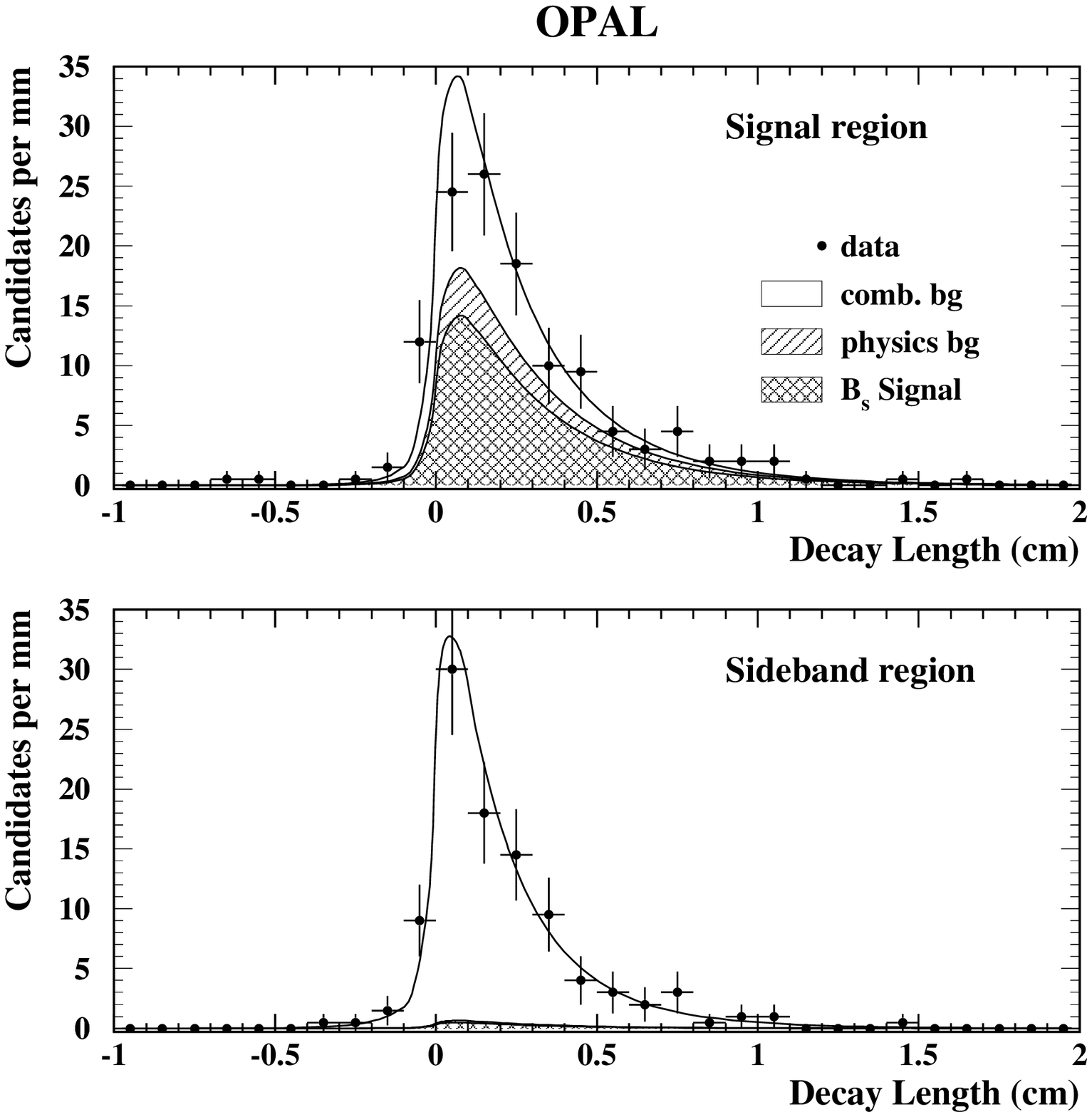}
\end{center}
\vspace{-3 cm}
\addFig{Fitted decay length distribution}
\Caption{ 
Comparison of expectation from fit results and actual decay length
distributions. The fit results are shown by the curves, with
individual components shaded, while the decay lengths measured in the
data are shown by the error bars. The sideband region was used in
the combinatorial background lifetime fit.
}
\label{fig:checklen}
\end{figure}

%       %       %        %       %       %       %       %       %       %
\label{sec:chedist}
The form of the likelihood of a candidate decaying with a measured decay
length $\li$ has been stated in Section~\ref{sec:like}. By setting the
various parameters to their nominal values and the amplitude to zero,
we acquired a prediction for the distribution of $\li$.
This prediction compared well with the actual
measured values of $\li$, as shown in Figure~\ref{fig:checklen}.

%       %       %        %       %       %       %       %       %       %
%\subsection{Results of lifetime fit}
%       %       %        %       %       %       %       %       %       %
The likelihood can also be used to measure the $\Bs$ lifetime by
%assuming very fast mixing (i.e. setting $\dms$ to a large value or
%equivalently by setting the amplitude to zero) and by 
ignoring the mixing tag and fitting the $\Bs$
lifetime to the data. The resulting lifetime is $1.57 \pm 0.17 \ps$, which
is consistent with the world average value of $1.54 \pm 0.07 \ps$.
\subsection{Oscillation fit on simulated events}
\begin{figure}[tb]
\centering
\epsfxsize=13cm
\begin{center}
    \leavevmode
    \epsffile{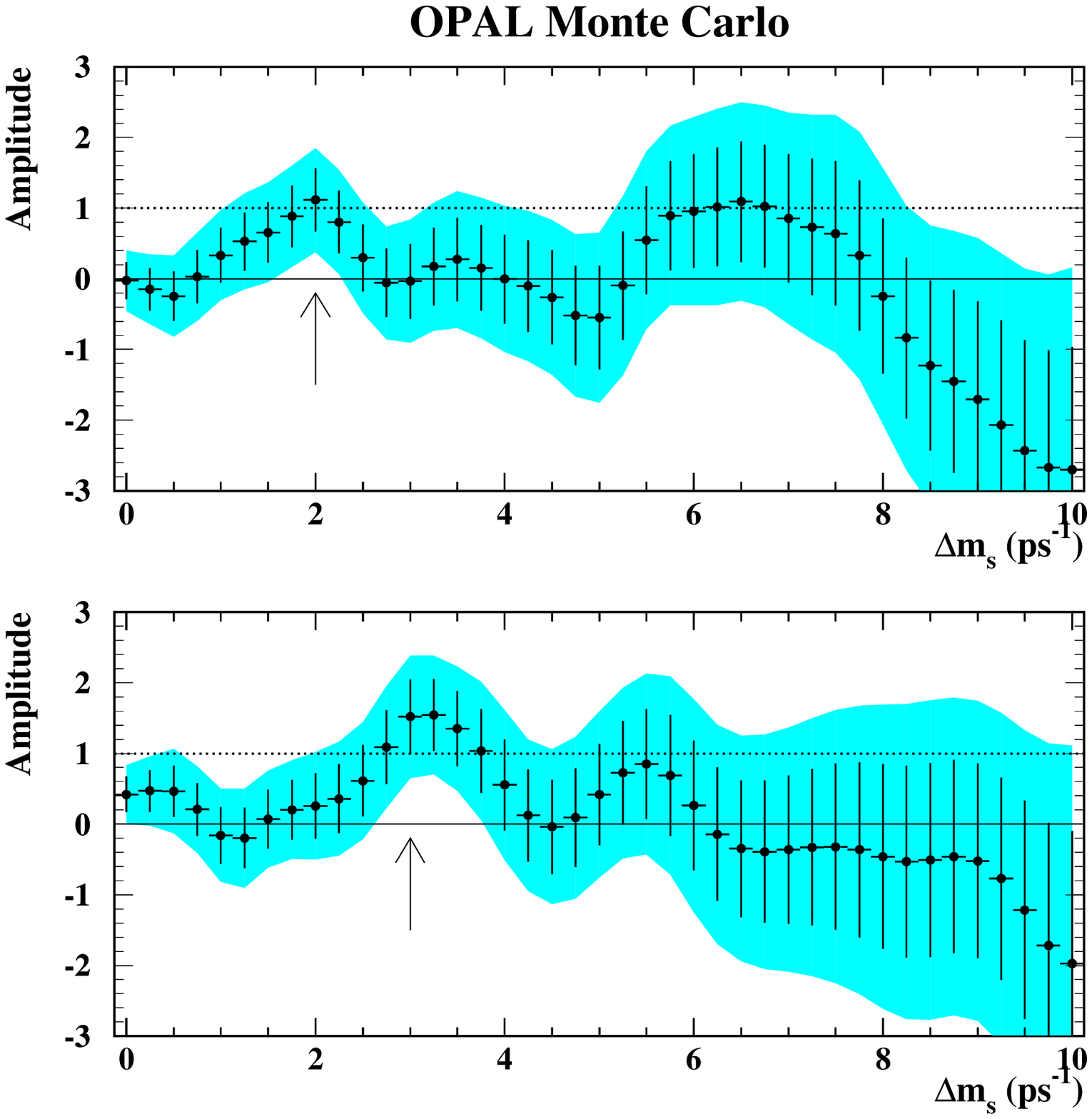}
\end{center}
\vspace{-5 mm}
\addFig{Results of amplitude fit for simulated events}
\Caption{ 
Measured $\Bs$ oscillation amplitude as a function of $\dms$ in
simulated events with a true $\dms$ of $2.0\psm$ in the upper plot,
and $3.0\psm$ in the lower plot.
The error bars represent the 1$\sigma$ statistical uncertainties.
The shaded bands show the $\pm 1.645 \sigma$ region (systematic
effects not included).
The arrows point at the simulated value of $\dms$.
}
\label{fig:ampmc}
\end{figure}

%       %       %        %       %       %       %       %       %       %
\label{sec:cheMC}
A likelihood fit for $\dms$ was performed on a simulated Monte Carlo event
sample having the same statistics and estimated composition as the
data, with an oscillation at a true frequency of either $\dms=2.0\psm$
or $\dms=3.0\psm$. The fit yielded $\dms=1.98^{+0.15}_{-0.14}\psm$ and 
$\dms=3.16^{+0.26}_{-0.23}\psm$ respectively, in agreement with the
input values.  
% updated 2.0,3.0 at 7/00
% note: peak at 0 in 3.0 simulation is from the mixcount. Without
% running the fitter we can estimate the amp. at zero from the mix
% count as described in ../text/mixCount. For this sample we get
% .45@0.8 while the fitted amp there is .42 ; so this is a
% stat. fluctuation and not a fitter artifact.               
% This two event samples are statistically correlated, because
% the same Monte Carlo events were used to simulate different
% oscillation frequencies.
% actually it's not ALL the same events, just some.
Performing an amplitude fit (ignoring systematic uncertainties)
on the same Monte Carlo events
yielded the results shown in Figure~\ref{fig:ampmc}. As expected, the
amplitude is consistent with $1$ at the true value of $\dms$.

The sensitivity of the analysis is defined as the expected highest
oscillation frequency excluded at the 95\% confidence level, given
that the true $\dms$ is infinitely high. 
% A similar likelihood fit including systematic effects was
% performed with a
% simulated oscillation frequency of $\dms=25.0~\psm$, which is well above
% our expected sensitivity and so equivalent to an infinitely fast oscillation.
Given an infinitely high $\dms$ the expectation value of the amplitude
at all fitted values of $\dms$ is zero~\cite{moser}.
This allows us to evaluate the sensitivity as the frequency at
which the resulting $1.645\sigma$ line rises above an amplitude of one,
as shown in Figure~\ref{fig:amp}.
The experimental sensitivity of this analysis is $\sensitivity\psm$.

%%%%%%%%%%%%%%%%%%%%%%%%%%%%%%%%%%%%%%%%%%%%%%%%%%%%%%%%%%%%%%%%%%%%%%%%
\section{Conclusion}
\begin{figure}[htbp]
\centering
\epsfxsize=17cm
\begin{center}
  \leavevmode
  \epsffile{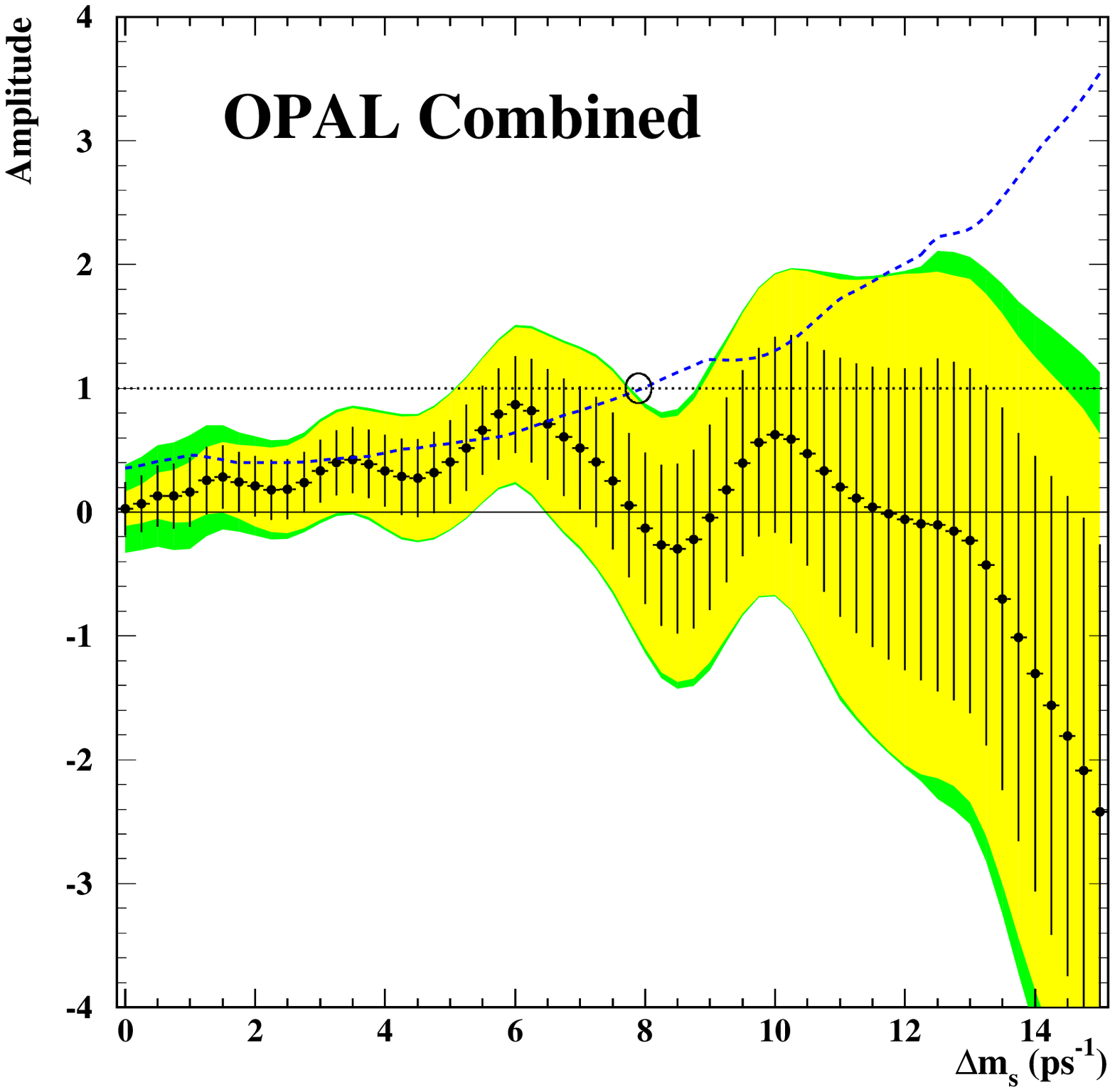}
\end{center}
\vspace{-5 mm}
\addFig{Combined results}
\Caption{ 
Measured $\Bs$ oscillation amplitude as a function of $\dms$ combined
for this analysis and the previous OPAL analysis.
The error bars represent the 1$\sigma$ statistical uncertainties.
The shaded bands show the $\pm 1.645 \sigma$ region, with and without
including systematic effects. The values of $\dms$ where the shaded
region lies below the ${\cal A}=1$ line are excluded \clnf.
The dashed line is 1.645$\sigma$ used to determine the
experimental sensitivity, which is indicated by the circle. 
}
\label{fig:combined}
\end{figure}

%%%%%%%%%%%%%%%%%%%%%%%%%%%%%%%%%%%%%%%%%%%%%%%%%%%%%%%%%%%%%%%%%%%%%%%%
\label{sec:conclusion}
A sample of $\Bs$ decays obtained using $\Dsl$ combinations was used
 to study $\Bs$ oscillation.
The estimated sensitivity of the analysis is $\sensitivity\psm$.
The resulting, stand-alone, lower limit from our analysis is
significantly lower, at $\limit\psm$. This limit is not competitive with
existing limits. However, this does not diminish the
contribution of this analysis to the world combined measurement, which
is at the rapid oscillation region. 
The previous OPAL measurement~\cite{martin} was an inclusive
 measurement, and so has only a negligable statistical
 correlation with our measurement. Its sensitivity was $6.7\psm$ and
 it set a lower limit of $5.1\psm$.
Combining it with our results 
 we get the
 combined measurement shown in Figure~\ref{fig:combined}, a
 sensitivity of $8.0\psm$ and a limit of $5.1\psm$.
% from combos fit on hpplus:/afs/cern.ch/user/a/aharel/public/dms/ftn10

%%%%%%%%%%%%%%%%%%%%%%%%%%%%%%%%%%%%%%%%%%%%%%%%%%%%%%%%%%%%%%%%%%%%%%%%%%
\section{Acknowledgements}
%\par
%Acknowledgements:
%\par

% \appendix
% \par
% Acknowledgements:
% \par
We particularly wish to thank the SL Division for the efficient operation
of the LEP accelerator at all energies
 and for their continuing close cooperation with
our experimental group.  We thank our colleagues from CEA, DAPNIA/SPP,
CE-Saclay for their efforts over the years on the time-of-flight and trigger
systems which we continue to use.  In addition to the support staff at our own
institutions we are pleased to acknowledge the  \\
Department of Energy, USA, \\
National Science Foundation, USA, \\
Particle Physics and Astronomy Research Council, UK, \\
Natural Sciences and Engineering Research Council, Canada, \\
Israel Science Foundation, administered by the Israel
Academy of Science and Humanities, \\
Minerva Gesellschaft, \\
Benoziyo Center for High Energy Physics,\\
Japanese Ministry of Education, Science and Culture (the
Monbusho) and a grant under the Monbusho International
Science Research Program,\\
Japanese Society for the Promotion of Science (JSPS),\\
German Israeli Bi-national Science Foundation (GIF), \\
Bundesministerium f\"ur Bildung und Forschung, Germany, \\
National Research Council of Canada, \\
Research Corporation, USA,\\
Hungarian Foundation for Scientific Research, OTKA T-029328, 
T023793 and OTKA F-023259.\\

%%%%%%%%%%%%%%%%%%%%%%%%%%%%%%%%%%%%%%%%%%%%%%%%%%%%%%%%%%%%%%%%%%%%%%%%%%

%%%%%%%%%%%%%%%%%%%%%%%%%%%%%%%%%%%%%%%%%%%%%%%%%%%%%%%%%%%%%%%%%%%%%%%%
% Bibliography
%%%%%%%%%%%%%%%%%%%%%%%%%%%%%%%%%%%%%%%%%%%%%%%%%%%%%%%%%%%%%%%%%%%%%%%%

% \begin{thebibliography}{99}
% \begin{thebibliography}{10}

\end{document}